\documentclass[12pt,a4paper]{article}
\usepackage[margin=2cm]{geometry}
\usepackage[english]{babel}
\usepackage[latin1]{inputenc}
\usepackage{amsmath, amsthm, amssymb}
\usepackage[pdftex]{graphicx}
\usepackage{amssymb}
\usepackage{latexsym}
\usepackage{amstext}
\usepackage{epstopdf}

\newcommand{\ad}{\mathbf{a_2}}

\newcommand{\adc}{\mathbf{a_2^c}}
\newcommand{\add}{\mathbf{a_2^{(2)}}}
\newcommand{\adf}{\mathbf{a_2^f}}
\newcommand{\adgd}{\mathbf{a_2^{(2)}}}
\newcommand{\adgu}{\mathbf{a_2^{(1)}}}

\newcommand{\adpm}{\mathbf{a_2^\pm}}
\newcommand{\adu}{\mathbf{a_2^{(1)}}}
\newcommand{\af}{\mathbf{a_5}}
\newcommand{\au}{a_1}
\newcommand{\aud}{a_1^\dagger}
\newcommand{\aun}{\mathbf{a_1}}
\newcommand{\aut}{a_1^T}

\newcommand{\bu}{b_1}
\newcommand{\bud}{b_1^\dagger}
\newcommand{\bun}{\mathbf{b_1}}
\newcommand{\but}{b_1^T}
\newcommand{\be}{\begin{equation}}
\newcommand{\bea}{\begin{equation}\begin{array}}
\newcommand{\beas}{\begin{equation*}\begin{array}}
\newcommand{\bef}{\begin{flalign}}
\newcommand{\befs}{\begin{flalign*}}
\newcommand{\bes}{\begin{equation*}}
\newcommand{\cc}{\mathbf{C}}
\newcommand{\cu}{c_1}
\newcommand{\cud}{c_1^\dagger}
\newcommand{\cut}{c_1^T}
\newcommand{\ee}{\end{equation}}
\newcommand{\eea}{\end{array}\end{equation}}
\newcommand{\eeas}{\end{array}\end{equation*}}
\newcommand{\eef}{\end{flalign}}
\newcommand{\eefs}{\end{flalign*}}
\newcommand{\ees}{\end{equation*}}
\newcommand{\ekm}{\varepsilon_k^-}
\newcommand{\ekme}{\mathbf e_k^-}
\newcommand{\esm}{E_{7 \frac{1}{2}}}
\newcommand{\ekmp}{\varepsilon_k^{\mp}}
\newcommand{\ekp}{\varepsilon_k^+}
\newcommand{\ekpe}{\mathbf e_k^+}
\newcommand{\ekpm}{\varepsilon_k^{\pm}}

\newcommand{\ejpm}{\varepsilon_j^\pm}
\newcommand{\eo}{\mathbf{e_8}}

\newcommand{\epp}{\varepsilon^+}

\newcommand{\es}{\mathbf{e_6}}
\newcommand{\est}{\mathbf{e_7}}

\newcommand{\fq}{\mathbf{f_4}}
\newcommand{\fqe}{\mathfrak{f}}

\newcommand{\gd}{\mathbf{g_2}}
\newcommand{\glt}{\mathbf{gl(3)}}

\newcommand{\gzd}{\mathbf{g_0^2}}
\newcommand{\hu}{\mathbf{H}}
\newcommand{\imi}{\mathbf{i}}
\newcommand{\jac}{\mathfrak{J}}
\newcommand{\jdot}{\!\cdot\!}
\newcommand{\jo}{\mathbf{J}}

\newcommand{\jotn}{\mathbf{J_3^n}}
\newcommand{\jobtn}{\mathbf{\overline J_3^{\raisebox{-2 pt}{\scriptsize \textbf n}}}}
\newcommand{\jotu}{\mathbf{J_3^1}}
\newcommand{\jobtu}{\mathbf{\overline J_3^{\raisebox{-2 pt}{\scriptsize \textbf 1}}}}
\newcommand{\jotd}{\mathbf{J_3^2}}
\newcommand{\jobtd}{\mathbf{\overline J_3^{\raisebox{-2 pt}{\scriptsize \textbf 2}}}}
\newcommand{\jotq}{\mathbf{J_3^4}}
\newcommand{\jobtq}{\mathbf{\overline J_3^{\raisebox{-2 pt}{\scriptsize \textbf 4}}}}
\newcommand{\joto}{\mathbf{J_3^8}}
\newcommand{\jobto}{\mathbf{\overline J_3^{\raisebox{-2 pt}{\scriptsize \textbf 8}}}}

\newcommand{\lk}{\mathfrak{L}}

\newcommand{\Lvxp}{L_{\mathbf x^+}}
\newcommand{\Lvxm}{L_{\mathbf x^-}}
\newcommand{\Lvxpm}{L_{\mathbf x^\pm}}

\newcommand{\Lvyp}{L_{\mathbf y^+}}
\newcommand{\Lvym}{L_{\mathbf y^-}}
\newcommand{\Lvypm}{L_{\mathbf y^\pm}}

\newcommand{\mep}{\star}
\newcommand{\Mu}{M^{(1)}}
\newcommand{\Md}{M^{(2)}}
\newcommand{\nbf}{\mathbf{n}}
\newcommand{\oo}{\textbf{\large $\mathfrak C$}}
\newcommand{\pu}{\phantom{=}}

\newcommand{\qq}{\mathbf{Q}}
\newcommand{\rep}{\mathbf \varrho}
\newcommand{\rom}{\rho^-}
\newcommand{\rmp}{\rho^\mp}
\newcommand{\rop}{\rho^+}
\newcommand{\rpm}{\rho^\pm}
\newcommand{\rr}{\mathbf{R}}

\newcommand{\sltc}{\mathbf{sl(3, \cc)}}
\newcommand{\sltq}{\mathbf{sl(3, \qq)}}
\newcommand{\sqd}{\frac{\sqrt2}{2}}
\newcommand{\sqs}{\frac{\sqrt6}{6}}
\newcommand{\sqst}{\frac{\sqrt6}{3}}
\newcommand{\str}{\text{str}}

\newcommand{\ud}{\mathbf{u_2}}
\newcommand{\uu}{\mathbf{u_1}}

\newcommand{\vx}{\mathbf x}
\newcommand{\vxi}{\boldsymbol \xi}
\newcommand{\vxim}{\boldsymbol \xi^-}
\newcommand{\vxip}{\boldsymbol \xi^+}
\newcommand{\vxm}{\mathbf x^-}
\newcommand{\vxp}{\mathbf x^+}
\newcommand{\vxpm}{\mathbf x^\pm}
\newcommand{\vy}{\mathbf y}
\newcommand{\vym}{\mathbf y^-}
\newcommand{\vyp}{\mathbf y^+}
\newcommand{\vypm}{\mathbf y^\pm}

\newcommand{\vz}{\mathbf z}
\newcommand{\vzz}{\boldsymbol \zeta}
\newcommand{\vzm}{\boldsymbol \zeta^-}
\newcommand{\vzp}{\boldsymbol \zeta^+}
\newcommand{\xs}{x^\#}

\numberwithin{equation}{section}
\begin{document}
%
\begin{titlepage}
\begin{center}

\vskip 5.5cm

{\huge{\bf Exceptional Lie Algebras,\\\vskip 0.5cm SU(3) and Jordan Pairs}}\\\vskip 0.5cm {\Large{\bf Part 2: Zorn-type Representations}}

\vskip 2.5cm

{\large{\bf Alessio Marrani\,$^1$ and  Piero Truini\,$^2$}}

\vskip 40pt

 {\it ${}^1$  Instituut voor Theoretische Fysica, KU Leuven,\\
Celestijnenlaan 200D, B-3001 Leuven, Belgium}\\\vskip 5pt
     \texttt{alessio.marrani@fys.kuleuven.be}

    \vspace{10pt}

{\it ${}^2$ Dipartimento di Fisica, Universit\` a degli Studi\\
via Dodecaneso 33, I-16146 Genova,  Italy}\\\vskip 5pt
\texttt{truini@ge.infn.it}

    \vspace{10pt}

\end{center}

\vskip 2.2cm

\begin{center} {\bf ABSTRACT}\\[3ex]\end{center}
A representation of the exceptional Lie algebras reflecting a simple unifying view, based on realizations in terms of Zorn-type matrices, is presented. The role of the underlying Jordan pair and Jordan algebra content is crucial in the development of the structure. Each algebra contains three Jordan pairs sharing the same Lie algebra of automorphisms and the same external $su(3)$ symmetry. Applications in physics are outlined.

\end{titlepage}

\newpage \setcounter{page}{1} \numberwithin{equation}{section}

\newpage\tableofcontents

\section{Introduction}

Groups in physics have at least a threefold valence. First, they
represent symmetries that, by definition, introduce elegance in all the
equations which are manifestly symmetry invariant. Moreover, symmetries also
arise as fundamental principles in constructing new theories, like, for
example, gauge symmetries for the Standard Model (SM) of particle physics,
conformal symmetry for string theory, or general covariance for the Einstein
theory of relativity. Finally, symmetries - hence groups - play a key role
in solving the equations of motion.

A particular class is represented by (semi)-simple Lie groups and
algebras, which find application in a large number of
mathematical and physical fields. All finite-dimensional complex Lie
algebras have been classified by Killing, whose proofs have been made
rigorous by Cartan, who has also extended the classification to
non-compact real forms. This classification has led to the discovery,
beyond the famous classical series, of five exceptional algebras
together with the corresponding real forms: $\mathbf{g}_{2}$, $\mathbf{f}%
_{4}$, $\mathbf{e}_{6}$, $\mathbf{e}_{7}$ and $\mathbf{e}_{8}$.

Exceptional Lie groups and algebras appear naturally as gauge symmetry
 groups of field theories which are low-energy limits of string
models \cite{Ramond}.

Various non-compact real forms of exceptional algebras occur in
supergravity theories in different dimensions as $U$-duality\footnote{%
Here $U$-duality is referred to as the \textquotedblleft
continuous\textquotedblright\ symmetries of \cite{CJ-1}. Their discrete
versions are the $U$-duality non-perturbative string theory symmetries
introduced in \cite{HT-1}.} . The related symmetric spaces are relevant by themselves for general relativity, because they are Einstein spaces \cite{Helgason}. In supergravity, some of these cosets, namely those  pertaining to the
non-compact real forms, are interpreted as scalar fields of the
associated non-linear sigma model (see \textit{e.g.} \cite{bt-1,bt-2}, and also \cite{LA08-review} for a review and list of Refs.). Moreover, they can
represent the charge orbits of electromagnetic fluxes of black holes when the Attractor Mechanism \cite{AM-Refs} is studied (\cite{BH-orbits}; for a
comprehensive review, see \textit{e.g.} \cite{small-orbits}), and they also appear as the \textit{moduli spaces} \cite{FM-2} for extremal black hole
attractors; this approach has been recently extended to all kinds of branes in supergravity \cite{brane orbits}. Fascinating group theoretical
structures arise clearly in the description of the Attractor Mechanism for
black holes in the Maxwell-Einstein supergravity , such as the so-called
magic exceptional $\mathcal{N}=2$ supergravity \cite{MESGT} in four
dimensions, which is related to the minimally non-compact real $\mathbf{e}%
_{7(-25)}$ form \cite{Dobrev-2} of $\mathbf{e}_{7}$.

The smallest exceptional Lie algebra, $\mathbf{g}_{2}$, occurs for instance
in the deconfinement phase transitions \cite{G2-1}, in random
matrix models \cite{G2-2}, and in matrix models related to $D$-brane physics
\cite{G2-3}; it also finds application to Montecarlo analysis \cite%
{MonteCarlo}.

$\mathbf{f}_{4}$ enters the construction of integrable models on exceptional
Lie groups and of the corresponding coset manifolds. Of particular interest,
from the mathematical point of view, is the coset manifold $\mathfrak{C}%
\mathbb{P}^{2}=F_{4}/Spin(9)$, the octonionic projective plane (see \textit{%
e.g.} \cite{baez}, and Refs. therein). Furthermore, the split real form $%
\mathbf{f}_{4(4)}$ has been recently proposed as the global symmetry of an
exotic ten-dimensional theory in the context of gauge/gravity correspondence
and \textquotedblleft magic pyramids" in \cite{ICL-Magic}.

Starting from the pioneering work of G\"{u}rsey \cite{gursey80,Gursey-2} on
Grand Unified theories (GUTs), exceptional Lie algebras have been related to
the study of the SM, and to the attempts to go beyond it: for example, the
discovery of neutrino oscillations, the fine tuning of the mixing matrices,
the hierarchy problem, the difficulty in including gravity, and so on. The
renormalization flow of the coupling constants suggests the unification of
gauge interactions at energies of the order of $10^{15}$ GeV, which can be
improved  and fine tuned by supersymmetry. In this framework the gauge group $G$
of Grand Unified theory (GUT) is expected to be simple, to contain the SM gauge group $%
SU(3)_{c}\times \ SU(2)_{L}\times \ U(1)_{Y}$ and also
to predict the correct spectra after spontaneous symmetry breaking. The
particular structure of the neutrino mixing matrix has led to the proposal
of $G$ given by the semi-direct product between the exceptional group $E_{6}$
and the discrete group $S_{4}$ \cite{E6-GUT}. For some mathematical studies on various real forms of $\mathbf{e}_{6}$, see e.g. \cite{Dobrev}, and Refs. therein.

Recently, $\mathbf{e}_{7}$ and \textquotedblleft groups of type $E_{7}$"
\cite{brown} have appeared in several indirectly related contexts. They have
been investigated in relation to minimal coupling of vectors and scalars in
cosmology and supergravity \cite{E7-cosmo}. They have been considered as
gauge and global symmetries in the so-called Freudenthal gauge theory \cite%
{FGT}. Another application is in the context of entanglement in quantum
information theory; this is actually related to its application to black
holes via the black-hole/qubit correspondence (see \cite{BH-qubit-Refs} for
reviews and list of Refs.). For various studies on the split real form of $\mathbf{e}_{7}$ and its application to maximal supergravity, see e.g. \cite{KS,KK,Bianchi-Ferrara,Brink}.

The largest finite-dimensional exceptional Lie algebra, namely $\mathbf{e}%
_{8}$, appears in supergravity \cite{MS} in its maximally non-compact
(split) real form, whereas the compact real form appears in heterotic string
theory \cite{GHMR}. Rather surprisingly, in recent times the popular press
has been dealing with $\mathbf{e}_{8}$ more than once. Firstly, the
computation of the Kazhdan-Lusztig-Vogan polynomials \cite{V} involved the
split real form of $\mathbf{e}_{8}$. Then, attempts at formulating a
\textquotedblleft theory of everything" were considered in \cite{L}, but
they were proved to be unsuccessful (\textit{cfr. e.g.} \cite{DG10}). More
interestingly, the compact real form of $\mathbf{e}_{8}$ appears in the
context of the cobalt niobate ($CoNb_{2}O_{6}$) experiment, making this the
first actual experiment to detect a phenomenon that could be modeled using $%
\mathbf{e}_{8}$ \cite{E8-exp}.

It should also be recalled that alternative approaches to quantum gravity,
such as loop quantum gravity, \cite{carlo} have also led towards the
exceptional algebras, and $\mathbf{e_{8}}$ in particular (see \textit{e.g.}
\cite{smoli}).

It is worth mentioning that the adjoint of $\mathbf{e}_{8}$ is its smallest fundamental representation; this sets $\mathbf{e}_{8}$ on a different
footing with respect to all other Lie algebras for unifying theories, which all exhibit a
fundamental representation of lower dimension than the adjoint - of
dimension $7$, $26$, $27$, $56$, in particular, for the exceptional algebras $\mathbf{g}_{2}$, $%
\mathbf{f}_{4}$, $\mathbf{e}_{6}$, $\mathbf{e}_{7}$ respectively. In
the framework of a unified physical theory, therefore, only an $\mathbf{e}_{8}$-based
model has matter particles, intermediate bosons, Higgs(es) \textit{etc.} all
in the same (adjoint, $248$-dimensional) representation.

There is a wide consensus in both mathematics and physics on the appeal of
the largest exceptional Lie algebra $\mathbf{e}_{8}$, considered by many
beautiful in spite of its complexity (for an explicit realization of its octic invariant, see \cite{Octic}).

\bigskip

\begin{figure}[tbp]
\begin{center}
\includegraphics{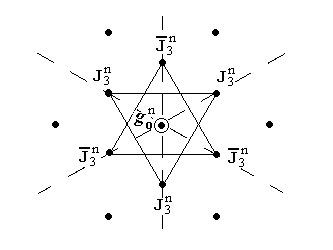}
\end{center}
\caption{A unifying view of the roots of exceptional Lie algebras}
\label{fig:diagram}
\end{figure}

The present paper is the continuation of a previous one \cite{pt1}, in which
the (finite-dimensional) exceptional Lie algebras were studied from a
unifying point of view represented by the diagram in figure \ref{fig:diagram}%
.

Figure \ref{fig:diagram} shows the projection of the roots of the
exceptional Lie algebras on a complex $\mathbf{su(3)}=\mathbf{a}_{2}$ plane,
recognizable by the dots forming the external hexagon, and it exhibits the
\textit{Jordan pair} content of each exceptional Lie algebra. There are
three Jordan pairs $(\jotn,\jobtn)$, each of which lies on an axis
symmetrically with respect to the center of the diagram. Each pair doubles
a simple Jordan algebra of rank $3$, $\jotn$, with involution - the
conjugate representation $\jobtn$, which is the algebra of $3\times 3$
Hermitian matrices over $\hu$, where $\hu=\rr,\,\cc  ,\,\qq,\,\oo$ for $\nbf=1,2,4,8$ respectively, stands for real, complex, quaternion, octonion algebras, the four composition
algebras according to Hurwitz's Theorem - see
\textit{e.g.} \cite{McCrimmon}.
Exceptional Lie algebras $\fq$, $\es$, $\est$, $\eo$ are obtained for $\nbf%
=1,2,4,8$, respectively. $\gd$ can be also represented in the same way, with
the Jordan algebra reduced to a single element; this corresponds to setting $%
\nbf=-2/3$; in Table \ref{grouptheorytable} below. The Jordan algebras $\mathbf{J}_{3}^{\mathbf{n}%
}$ (and their conjugate $\overline{\mathbf{J}_{3}^{\mathbf{n}}}$) globally
behave like a $\mathbf{3}$ (and a $\mathbf{\overline{3}}$) dimensional
representation of the outer $\mathbf{a}_{2}$. The algebra denoted by $\mathbf{%
g_{0}^{n}}$ in the center (plus the Cartan generator associated with the
axis along which the pair lies) is the algebra of the
automorphism group of the Jordan Pair; namely, $\mathbf{g_{0}^{n}}$ is the
the \textit{reduced} structure group of the corresponding
Jordan algebra $\mathbf{J}_{3}^{\mathbf{n}}$: $\mathbf{g_{0}^{n}=str}%
_0\left( \mathbf{J}_{3}^{\mathbf{q}}\right) $. Notice that $\mathbf{J}%
_{3}^{\mathbf{n}}$ fits into a $\left( 3\mathbf{n}+3\right) $-dimensional
irreducible representation of $\mathbf{g_{0}^{n}}$ itself.

The \textit{base field} considered throughout the present paper is $\mathbf{C%
}$. Therefore, all parameters in the whole paper are \textit{complex}
numbers. For instance, \textbf{J}$_{3}^{1}$ is a \textit{real} Jordan
algebra over the complex numbers, which means that the Hermitian conjugation
is the transposed of matrices: \textbf{J}$_{3}^{1}$ is an algebra of
symmetric complex matrices.

The reason for choosing complex numbers as base field lies in the fact that
we are dealing with the root diagrams of the Lie algebras, therefore we need
an algebraically closed field. The various real compact and non-compact forms
of the exceptional Lie algebras follow as a consequence, using some more or
less laborious tricks, whose treatment we leave to a future study, and they
do not affect the essential structure.

The \textit{real}, \textit{complex}, \textit{quaternion}, \textit{octonion}
attributes corresponding to setting $\mathbf{n}=1$, $2$, $4$, $8$ in \textbf{%
J}$_{3}^{\mathbf{n}}$ refer to algebras - over the complex field - whose
role is that of a book-keeping device: they are used in order to make the
language easier and more compact. In this sense, they fall naturally into
the Lie structure.

While the algebras $\mathbf{R}$ and $\mathbf{C}$ are \textit{commutative}, the algebras $\mathbf{Q}$ and $\mathfrak{C}$ are \textit{non-commutative} (octonions - \textit{Cayley numbers} - $%
\mathfrak{C}$ are also \textit{non-associative}), but they all are \textit{%
alternative}, a fundamental property without which our whole construction would
fall apart\footnote{%
\textit{Non-alternative} extensions beyond
$\mathfrak{C}$, such as \textit{sedenions} and \textit{trigintaduonions} (%
\textit{cfr.} \textit{e.g.} \cite{beyond-O}) would require a
different approach.}. They, however, have nothing to do with the base field
- the complex field $\mathbf{C}$ - of the corresponding Lie algebras. On the
other hand, it is true the opposite : having complex alternative algebras
allows to have nilpotents, which are as useful as $J^{+}$ and $J^{-}$ are in
the algebra of spin, or as creation and annihilation operators are in the
description of the quantum harmonic oscillator or in quantum field theory.

By varying $\mathbf{n}$, figure \ref{fig:diagram} depicts the following
decomposition, \cite{pt1} :%
\begin{equation}
\mathbf L^{\mathbf{n}}=\ad\oplus \mathbf{str}_0\left( \mathbf{J}_{3}^{\mathbf{n}}\right) \oplus \mathbf{3}\times
\mathbf{J}_{3}^{\mathbf{n}}\oplus \overline{\mathbf{3}}\times \overline{\mathbf{J}_{3}^{\mathbf{n}}},  \label{2}
\end{equation}
with the corresponding compact cases given in Table \ref{grouptheorytable}:\\
\begin{table}[h!]
\begin{center}
\begin{tabular}{|c||c|c|c|c|c|c|c|}
\hline
$\mathbf{n}$ & $8$ & $4$ & $2$ & $1$ & $0$ & $-2/3$ & $-1$ \\ \hline
\rule[-1mm]{0mm}{6mm} $\mathbf L^{\mathbf{n}}$ & $\mathbf{e}_{8\left(
-248\right) }$ & $\mathbf{e}_{7\left( -133\right) }$ & $\mathbf{e}_{6\left(
-78\right) }$ & $\mathbf{f}_{4\left( -52\right) }$ & $\mathbf{so}\left(
8\right) $ & $\mathbf{g}_{2\left( -14\right) }$ & $\mathbf{su}\left(
3\right) $ \\ \hline
\rule[-1mm]{0mm}{6mm} $\mathbf{str}_{0,c}$ & $\mathbf{e}_{6\left( -78\right)
}$ & $\mathbf{su}\left( 6\right) $ & $\mathbf{su}\left( 3\right) \oplus
\mathbf{su}\left( 3\right) $ & $\mathbf{su}\left( 3\right) $ & $\mathbf{u}%
\left( 1\right) \oplus \mathbf{u}\left( 1\right) $ & $-$ & $-$ \\ \hline
\end{tabular}%
\end{center}

\caption{The exceptional sequence \label{grouptheorytable}}
\end{table}

The sequence $\mathbf{L}^{\mathbf{n}}$ is usually named \textit{\textquotedblleft exceptional sequence"%
} (or \textit{\textquotedblleft exceptional series"}; see \textit{e.g.} \cite%
{LM-1}, and Refs. therein). This can be either interpreted as a sequence of
Lie algebras over the complex numbers $\mathbf{C}$, as we will consider
throughout the present investigation, or as a sequence of corresponding \textit{compact}
real forms.

It is here worth pointing out that, by considering suitable non-compact,
real forms, one obtains the $\mathbf{n}$-parametrized sequence of $U$%
-duality Lie algebras $\mathbf{L}^{\mathbf{n}}$ in $D=3$ (Lorentzian)
space-time dimensions\footnote{%
Jordan pairs of \textit{semi-simple} Euclidean Jordan
algebras of rank $3$ in supergravity theories (among which the
case of $\mathbf{so}\left( 8\right) $, $\mathbf{n}=0$) has been
presented in \cite{Jordan-Pairs-D=5}.\medskip} \cite{Jordan-Pairs-D=5} :%
\begin{equation}
\mathbf{L}^{\mathbf{n}}=\mathbf{sl}\left( 3,\mathbf{R}\right) \oplus \mathbf{%
str}_{0}\left( \mathbf{J}_{3}^{\mathbf{n}}\right) \oplus \mathbf{3}\times
\mathbf{J}_{3}^{\mathbf{n}}\oplus \mathbf{3}^{\prime }\times \mathbf{J}_{3}^{%
\mathbf{n}\prime }.  \label{2-bis}
\end{equation}%
Note that the reduced structure Lie algebra $\mathbf{str}_{0}\left( \mathbf{J%
}_{3}^{\mathbf{n}}\right) $, which, as stated above, is a suitable
non-compact real form of $\mathbf{g_{0}^{n}}$, is nothing but the $D=5$ $U$%
-duality Lie algebra. Also,
\begin{equation}
\mathbf{L}^{\mathbf{n}}=\mathbf{qconf}\left( \mathbf{J}_{3}^{\mathbf{n}%
}\right)  \label{qconf}
\end{equation}%
is the \textit{quasi-conformal} Lie algebra of $\mathbf{J}_{3}^{\mathbf{n}}$
\cite{GKN,GP-04}, \textit{i.e.} the $U$-duality Lie algebra in $D=3$ (see
\textit{e.g.} \cite{G-Lects} and \cite{small-orbits} for an introduction to
the application of Jordan algebras and their symmetries in supergravity%
\footnote{%
In these theories, the $U$-duality Lie algebra in $D=4$ (Lorentzian)
space-time dimensions is given by the \textit{conformal} Lie algebra $%
\mathbf{conf}\left( \mathbf{J}_{3}^{\mathbf{n}}\right) =\mathbf{aut}\left(
\mathfrak{F}\left( \mathbf{J}_{3}^{\mathbf{n}}\right) \right) $, where $%
\mathfrak{F}\left( \mathbf{J}_{3}^{\mathbf{n}}\right) $ denotes the
Freudenthal triple system constructed over $\mathbf{J}_{3}^{\mathbf{n}}$.},
and lists of Refs.). Suitable real, non-compact forms of all
exceptional Lie algebras can thus be characterized as \textit{quasi-conformal}
algebras\footnote{%
The case $\mathbf{n}=-1$ is trivial, and it corresponds to \textit{%
\textquotedblleft pure"} $\mathcal{N}=2$ supergravity in four-dimensional
Lorentzian space-time; therefore, it does not admit an uplift to five
dimensions, and it will henceforth not be considered. Moreover, $\mathbf{su}%
(2)$ might be considered as the $\mathbf{n}=-4/3$ element of the sequence in
Table below \eqref{2}, as well. However, this is a limit case of the \textquotedblleft
exceptional" sequence reported in Table \ref{grouptheorytable}, not pertaining to Jordan pairs
nor to supergravity in $D=3$ dimensions, and thus we will disregard it.} of
Euclidean simple Jordan algebras of rank $3$.

At group level, the algebraic decompositions (\ref{2}) and (\ref{2-bis}) are
Cartan decompositions respectively pertaining to the following \textit{%
maximal non-symmetric} embeddings:%
\begin{eqnarray}
QConf_{c}\left( \mathfrak{J}_{3}^{q}\right) &\supset &SU\left( 3\right)
\times Str_{0,c}\left( \mathfrak{J}_{3}^{q}\right) ; \\
QConf\left( \mathfrak{J}_{3}^{q}\right) &\supset &SL\left( 3,\mathbf{R}%
\right) \times Str_{0}\left( \mathfrak{J}_{3}^{q}\right) .
\end{eqnarray}%
As mentioned above, the non-semi-simple part of the r.h.s. of (\ref{2}) and (%
\ref{2-bis}) is given by a triplet of Jordan pairs.

Finally, we recall that in \cite{Jordan-Pairs-D=5}, by exploiting the Jordan
pair structure of $U$-duality Lie algebras in $D=3$ and the relation to the
\textit{super-Ehlers} symmetry in $D=5$ \cite{super-Ehlers-1}, the massless
multiplet structure of the spectrum of a broad class of $D=5$ supergravity
theories was investigated.\bigskip

In general, many properties of Lie algebras and groups can be already
inferred from abstract theoretical considerations; however, for most
applications, it is useful to have explicit concrete
realizations in terms of matrices\footnote{%
Explicit realizations of exceptional groups have been obtained \textit{e.g.}
in \cite{Exc-papers}. Our results, however, displays a much more manageable
form, with manifest $\ad$ covariance, as a consequence of the
full exploitation of the underlying Jordan pair structure.}.

In this paper we develop the results of \cite{pt1} and fully exploit
Jordan pairs and the corresponding unifying view depicted in figure \ref%
{fig:diagram}. We introduce \textit{Zorn-type} matrix realizations of all
exceptional finite-dimensional Lie algebras, which make
the Jordan pair structure manifest and are written in the form of a $2\times
2$ matrix, endowed with a quite peculiar matrix product accounting for the
complexity and non-associativity of the underlying structure. As a
consequence of (\ref{qconf}), this corresponds to the explicit construction
of \textit{Zorn-type} matrix realizations of the compact form of
quasi-conformal algebras of simple Jordan algebras of rank $3$; we
point out that in the present paper we will deal with Lie algebras over $%
\mathbb{C}$, leaving the analysis of real forms to future investigation.\\

The paper is organized as follows.

In section \ref{sec:jp} we briefly review the concept of a Jordan pair. Most
of the section can be found also in \cite{pt1} and is repeated here for
completeness.

For the same reason, as well as for introducing some notation, we present in
section \ref{sec:octonions} a summary on the octonion algebra and its
representation through the Zorn matrices, on which we base the development
of our representations. The key idea which we exploit here is that the
octonions' non-associativity can be cast into a properly defined product of $%
2\times 2$ complex matrices.

With this in mind, we are able to define, formally using $2\times 2$
matrices, a representation of $\gd$ in section \ref{sec:g2}, $\fq$ in
section \ref{sec:f4} (where we also make a comparison with Tits'
construction), $\es$ in section \ref{sec:e6}, $\est$ in section \ref{sec:est}%
. In section \ref{sec:jacobi} we prove the Jacobi identity for all these
algebras.

In the case of $\eo$, section \ref{sec:e8}, a new difficulty occurs due to
non-associativity. Not only the octonions are non-associative, but so is the
underlying standard matrix product of the Jordan algebra elements. This
forces a new definition of matrix elements and of their product, which still
allows us to formally describe the representation of $\eo$ through $2\times
2 $ matrices. The proof of the Jacobi identity for this case heavily relies
on the Jordan Pair axioms, and it is presented in section \ref{sec:jacobieo}.

The paper ends with some proposals of future developments of the present
work.

\section{Jordan Pairs}\label{sec:jp} In this section we review the concept of a Jordan Pair, \cite{loos1} (see also \cite{McCrimmon} for an enlightening overview).

Jordan Algebras have traveled a long journey, since their appearance in the 30's \cite{jvw}. The modern formulation \cite{jacob1} involves a quadratic map $U_x y$ (like $xyx$ for associative algebras) instead of the original symmetric product $x \jdot y = \frac12(xy + yx)$. The quadratic map and its linearization $V_{x,y} z = (U_{x+z} - U_x - U_z)y$ (like $xyz+zyx$ in the associative case) reveal  the mathematical structure of Jordan Algebras much more clearly, through the notion of inverse, inner ideal, generic norm, \textit{etc}. The axioms are:
\begin{equation}
U_1 = Id \quad , \qquad
U_x V_{y,x} = V_{x,y} U_x \quad  , \qquad
U_{U_x y} = U_x U_y U_x
\label{qja}
\end{equation}
The quadratic formulation led to the concept of Jordan Triple systems \cite{myb}, an example of which is a pair of modules represented by rectangular matrices. There is no way of multiplying two matrices $x$ and $y$ , say $n\times m$ and $m\times n$ respectively, by means of a bilinear product. But one can do it using a product like $xyx$, quadratic in $x$ and linear in $y$. Notice that, like in the case of rectangular matrices, there needs not be a unity in these structures. The axioms are in this case:
\begin{equation}
U_x V_{y,x} = V_{x,y} U_x \quad  , \qquad
V_{U_x y , y} = V_{x , U_y x} \quad , \qquad
U_{U_x y} = U_x U_y U_x
\label{jts}
\end{equation}

Finally, a Jordan Pair is defined just as a pair of modules $(V^+, V^-)$ acting on each other (but not on themselves) like a Jordan Triple:
\begin{equation}\begin{array}{ll}
U_{x^\sigma} V_{y^{-\sigma},x^\sigma} &= V_{x^\sigma,y^{-\sigma}} U_{x^\sigma}
\\
V_{U_{x^\sigma} y^{-\sigma} , y^{-\sigma}} &= V_{x ^\sigma, U_{y^{-\sigma}} x^\sigma} \\
U_{U_{x^\sigma} y^{-\sigma}} &= U_{x^\sigma} U_{y^{-\sigma}} U_{x^\sigma}\end{array}
\label{jp}
\end{equation}
where $\sigma = \pm$ and $x^\sigma \in V^{+\sigma} \, , \; y^{-\sigma} \in V^{-\sigma}$.

Jordan pairs are strongly related to the Tits-Kantor-Koecher construction of Lie Algebras $\lk$ \cite{tits1}-\nocite{kantor1}\cite{koecher1} (see also the interesting relation to Hopf algebras, \cite{faulk}):
\begin{equation}
\lk = J \oplus \str(J) \oplus \bar{J} \label{tkk}
\end{equation}
where $J$ is a Jordan algebra and $\str(J)= L(J) \oplus Der(J)$ is the structure algebra of $J$ \cite{McCrimmon}; $L(x)$ is the left multiplication in $J$: $L(x) y = x \jdot y$ and $Der(J) = [L(J), L(J)]$ is the algebra of derivations of $J$ (the algebra of the automorphism group of $J$) \cite{schafer1}\cite{schafer2}.

 In the case of complex exceptional Lie algebras, this construction applies to $\est$, with $J = \joto$, the 27-dimensional exceptional Jordan algebra of $3 \times 3$ Hermitian matrices over the complex octonions, and $\str(J) = \es \otimes \cc$ - $\mathbf{C}$ denoting the complex field. The algebra $\es$ is called the \emph{reduced structure algebra} of $J$, $\str_0(J)$, namely the structure algebra with the generator corresponding to the multiplication by a complex number taken away: $\es = L(J_0) \oplus Der(J)$, with $J_0$ denoting the traceless elements of $J$.

We conclude this introductory section with some standard definitions and identities in the theory of Jordan algebras and Jordan pairs, with particular reference to $\jotn, \mathbf{n}= 1,2,4,8$.
If $x,y \in \jotn$ and $xy$ denotes their standard matrix product, we denote by $x\jdot y := \frac12 (xy + yx)$ the Jordan product of $x$ and $y$. The Jordan identity is the power associativity with respect to this product:
\be\label{pass}
x^2 \jdot (x\jdot z) - x \jdot (x^2 \jdot z) = 0,
\ee

Another  fundamental product is the {\it sharp} product $\#$, \cite{McCrimmon}. It is the linearization of $\xs := x^2 - t(x) x - \frac12(t(x^2) - t(x)^2)I$, with $t(x)$ denoting the trace of $x\in \jotn$, in terms of which we may write the fundamental cubic identity for $\jotn, \mathbf{n}= 1,2,4,8$:
\be\label{cubic}
\xs\jdot  x = \frac13 t(\xs\!, x) I \quad \text{or} \quad x^3 - t(x) x^2 + t(\xs) x - \frac13 t(\xs\! , x) I = 0
\ee
where we use the notation $t(x,y) := t(x\jdot y)$ and  $x^3 = x^2 \jdot x$ (notice that for $\joto$, because of non-associativity, $x^2 x \ne x x^2$ in general).

The triple product is defined as, \cite{McCrimmon}:
\bea{ll}\label{vid}
\{ x , y , z \} := V_{x,y}z :&= t(x,y) z + t(z,y) x - (x \# z) \# y \\
&= 2 \left[ (x \jdot y)\jdot z +  (y \jdot z)\jdot x - (z \jdot x)\jdot y \right]
\eea

Notice that the last equality of \eqref{vid} is not trivial at all. $V_{x,y}z$ is the linearization of the quadratic map $U_xy$. The equation (2.3.15) at page 484 of \cite{McCrimmon} shows that:
\be\label{uid}
U_x y = t(x,y) x  - x ^\# \# y = 2 (x \jdot y)\jdot x - x^2 \jdot y
\ee

We shall make use of the following identities, which can be derived from the Jordan Pair axioms, \cite{loos1}:
\be
\left[ V_{x,y} , V_{z,w} \right] = V_{{V_{x,y} z},w} - V_{z,{V_{x,y} w}}
\label{comv}
\ee
and, for $D = (D_+,D_-)$ a derivation of the Jordan Pair $V$ and $\beta(x,y) = (V_{x,y}, - V_{y,x})$,
\be
[D, \beta(x,y)] = \beta (D_+(x),y) + \beta(x, D_-(y))
\label{dib}
\ee


\section{Octonions}\label{sec:octonions}

\begin{figure}
\begin{center}
\includegraphics{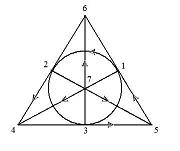}
\caption{Fano diagram for the octonions' products\label{fig:fano}}
\end{center}
\end{figure}
As we introduced in Sec. 1, $\oo$ stands for the algebra of the octonions (Cayley numbers) over the complex field  $\cc$ whose multiplication rule goes according to the Fano diagram in figure \ref{fig:fano} (for earlier studies, see e.g. \cite{Dund}).

If $a \in \oo$ we write $a = a_0 + \sum_{k=1}^7{a_k u_k}$, where $a_k \in \cc$ for $k = 1, \dots , 7$ and $u_k$ for $k = 1, \dots , 7$ denote the octonion imaginary units. We denote by $i$ the the imaginary unit in $\cc$.

Thence, we introduce 2 idempotent elements:
$$\rpm = \frac{1}{2}(1 \pm i u_7) $$
and 6 nilpotent elements:
$$\ekpm = \rpm u_k \ , \quad k = 1,2,3 $$
One can readily check that:
\begin{equation}
\begin{array}{ll}
 & (\rpm)^2 = \rpm \quad , \quad \rpm \rmp = 0 \\ \\
 & \rpm \ekpm = \ekpm \rmp = \ekpm \\ \\
 & \rmp \ekpm = \ekpm \rpm = 0 \\ \\
 & (\ekpm)^2 = 0 \quad , \qquad k = 1,2,3 \\ \\
& \ekpm \varepsilon_{k+1}^\pm = - \varepsilon_{k+1}^\pm \ekpm  = \varepsilon_{k+2}^\mp \qquad \text{(indices modulo 3)} \\ \\
& \ejpm \ekmp = 0 \qquad j \ne k \\ \\
& \ekpm \ekmp = - \rpm \quad , \qquad k = 1,2,3
\end{array}\end{equation}

It is known that octonions can be represented by Zorn matrices, \cite{zorn}.
If $a \in\oo$ , $A^\pm \in \cc^3$ is a vector with complex components $\alpha_k^{\pm}$ , $k=1,2,3$ (and we use the
standard summation convention over repeated indices throughout), then we have the identification:
\begin{equation}
a = \alpha_0^+ \rho^+ +\alpha_0^- \rho^- + \alpha_k^+
\varepsilon_k^+ +\alpha_k^- \varepsilon_k^- \longleftrightarrow
\left[
\begin{array}{cc}
\alpha_0^+ &  A^+ \\
A^- & \alpha_0^-
\end{array} \right];
\label{oct} \end{equation}
therefore, through Eq. \eqref{oct}, the product of $a, b \in \mathfrak{C}$ corresponds to:
\begin{equation}
\begin{array}{c}
\left [ \begin{array}{cc} \alpha^+ & A^+ \\ A^- & \alpha^-
\end{array}\right]
\left [ \begin{array}{cc} \beta^+ & B^+ \\ B^- & \beta^-
\end{array}\right] \\ =
\left [\begin{array}{cc} \alpha^+ \beta^+ + A^+ \cdot B^- &
\alpha^+ B^+ + \beta^- A^+ + A^- \wedge B^- \\
\alpha^- B^- + \beta^+ A^- + A^+ \wedge B^+ & \alpha^- \beta^- +
A^- \cdot B^+ \end{array} \right],
\label{zorn1}
\end{array}
\end{equation}
where $A^\pm \cdot B^\mp = - \alpha_K^\pm \beta_k^\mp$ and $A
\wedge B$ is the standard vector product of $A$ and $B$.

\section{$\gd$ action on Zorn matrices}\label{sec:g2}

In this section, we derive the matrix representation of $\gd$, and its action on Zorn matrices.  Let $a, b, c \in \oo$. Then the derivations of the octonions, \cite{schafer1} \cite{loos2}, can be written as $D_{a,b}$:
$$D_{a,b} c = \frac13 [[a,b],c] - (a,b,c) \qquad \text{ where } \ (a,b,c) = (ab)c - a(bc)$$

We choose the following $\gd$ generators, for $k = 1,2,3$ (mod $3$):
\begin{flalign*}
d^{\pm}_k &= \mp\  D_{\varepsilon_{k+1}^\pm,\varepsilon_{k+2}^\mp} = \mp\ L_{\varepsilon_{k+1}^{\pm}} L_{\varepsilon_{k+2}^{\mp}}\\
H_1 &= \tfrac {\sqrt{2}}2 \left(  D_{\varepsilon_1^- ,\varepsilon_1^+} - D_{\varepsilon_2^- ,\varepsilon_2^+} \right) = \tfrac{\sqrt2}2\left(  L_{\varepsilon_1^-} L_{\varepsilon_1^+} - L_{\varepsilon_2^-} L_{\varepsilon_2^+} \right)\\
H_2 &= \tfrac{\sqrt{6}}6 \left(  D_{\varepsilon_1^- ,\varepsilon_1^+} + D_{\varepsilon_2^- ,\varepsilon_2^+} - 2  D_{\varepsilon_3^- ,\varepsilon_3^+} \right) \\
&= \tfrac{\sqrt{6}}6 \left(  L_{\varepsilon_1^-} L_{\varepsilon_1^+} + L_{\varepsilon_2^-} L_{\varepsilon_2^+} - 2 L_{\varepsilon_3^-} L_{\varepsilon_3^+}\right)\\
g^{\pm}_k &=  3\ D_{\rpm,\ekpm} = L_{\ekpm} - R_{\ekpm}  - 3 L_{\rmp} L_{\ekpm}
\end{flalign*}

We notice that $ D_{\rho^+,\rho^-} = 0$,  $ D_{\rho^+,\varepsilon_k^\pm} = - D_{\rho^-,\varepsilon_k^\pm} = \mp D_{\varepsilon_{k+1}^\mp ,\varepsilon_{k+2}^\mp}$ and that $ D_{\varepsilon_1^- ,\varepsilon_1^+} + D_{\varepsilon_2^- ,\varepsilon_2^+} +  D_{\varepsilon_3^- ,\varepsilon_3^+} = 0$, hence the 14 generators introduced above span all the derivations of $\oo$.

The action of these generators on $a\in \oo$, $a = \alpha_0^+ \rho^+ +\alpha_0^- \rho^- + \alpha_k^+
\varepsilon_k^+ +\alpha_k^- \varepsilon_k^-$ is:
\begin{flalign*}
d_k^\pm \ &:\  a \rightarrow \pm (\alpha_{k+2}^\pm \varepsilon_{k+1}^\pm - \alpha_{k+1}^\mp \varepsilon_{k+2}^\mp)\\
H_1 \ &:\   a  \rightarrow \tfrac{\sqrt{2}}2 ( \alpha_1^+ \varepsilon_1^+ -
 \alpha_2^+ \varepsilon_2^+ - \alpha_1^- \varepsilon_1^- + \alpha_2^- \varepsilon_2^- )\\
H_2 \ &:\   a  \rightarrow \tfrac{\sqrt{6}}6 ( \alpha_1^+ \varepsilon_1^+ +
 \alpha_2^+ \varepsilon_2^+ - 2 \alpha_3^+ \varepsilon_3^+ - \alpha_1^- \varepsilon_1^- - \alpha_2^- \varepsilon_2^- + 2 \alpha_3^- \varepsilon_3^- )\\
g_k^\pm \ &:\  a \rightarrow  - \alpha_{k+1}^\pm \varepsilon_{k+2}^\mp +
 \alpha_{k+2}^\pm \varepsilon_{k+1}^\mp -  \alpha_k^\mp (\rho^\pm - \rho^\mp) - ( \alpha_0^\pm  -  \alpha_0^\mp) \varepsilon_k^\pm
\end{flalign*}

One can thus readily check that $[H_1,H_2]  =0$ and that the $g_k^\pm$'s are eigenvectors of $(H_1,H_2)$, with respect to the Lie product, with eigenvalues $(\pm \tfrac{\sqrt{2}}2, \pm \tfrac{\sqrt{6}}6)$  and $(0, \pm \tfrac{\sqrt{6}}3)$; the same with for $d_k^\pm$'s with eigenvalues $(\pm \tfrac{\sqrt{2}}2, \pm \tfrac{\sqrt{6}}2)$  and $(\pm {\scriptstyle \sqrt{2}},0)$, namely :
\begin{flalign*}
[H_1,H_2] &= 0 \\
[H_1, g_1^\pm] &= \pm \tfrac{\sqrt2}2 g_1^\pm &[H_2, g_1^\pm] &= \pm \tfrac{\sqrt6}6 g_1^\pm &[H_1, d_1^\pm] &= \mp \tfrac{\sqrt2}2 d_1^\pm &[H_2, d_1^\pm] &= \pm \tfrac{\sqrt6}2 d_1^\pm \\
[H_1, g_2^\pm] &= \mp \tfrac{\sqrt2}2 g_2^\pm &[H_2, g_2^\pm] &= \pm \tfrac{\sqrt6}6 g_2^\pm &[H_1, d_2^\pm] &= \mp \tfrac{\sqrt2}2 d_2^\pm &[H_2, d_2^\pm] &= \mp \tfrac{\sqrt6}2 d_2^\pm \\
[H_1, g_3^\pm] &= 0 &[H_2, g_3^\pm] &= \mp \tfrac{\sqrt6}3 g_3^\pm &[H_1, d_3^\pm] &= \pm {\scriptstyle \sqrt2}\ d_3^\pm &[H_2, d_3^\pm] &= 0
\end{flalign*}

Therefore, we have found out that the $d_k^\pm$ generators correspond to the external $\ad$ in the root diagram of $\gd$, whereas the $g_k^\pm$ generators correspond to the internal hexagon ($3$ and $\bar 3$ of $\ad$).

The remaining non-vanishing commutation relations are:
\begin{flalign*}
&[d_k^\pm ,d_{k+1}^\pm] = \pm d_{k+2}^\mp \\
&[d_1^+,d_1^-] =- \tfrac12 ({\scriptstyle \sqrt2} H_1 - {\scriptstyle \sqrt6} H_2) \quad [d_2^+,d_2^-] =- \tfrac12 ({\scriptstyle \sqrt2} H_1 + {\scriptstyle \sqrt6} H_2) \quad [d_3^+,d_3^-] ={\scriptstyle \sqrt2} H_1 \\
&[d_k^\pm ,g_{k+1}^\mp] = \mp g_{k+2}^\mp \quad \  [d_{k+1}^\pm ,g_k^\pm] = \pm g_{k+2}^\pm  \\
&[g_k^\pm ,g_{k+1}^\mp] = \mp 3 d_{k+2}^\pm \quad [g_k^\pm ,g_{k+1}^\pm] =  2 g_{k+2}^\mp  \\
&[g_1^+,g_1^-] =- \tfrac12 ({\scriptstyle 3 \sqrt2} H_1 + {\scriptstyle \sqrt6} H_2) \quad [g_2^+,g_2^-] =\tfrac12 ({\scriptstyle 3 \sqrt2} H_1 - {\scriptstyle \sqrt6} H_2) \quad [g_3^+,g_3^-] ={\scriptstyle \sqrt6} H_2 \\
\end{flalign*}

we now introduce the following complex algebra of $4\times 4$ Zorn-type matrices:
\begin{equation}
\left[
\begin{array}{cc}
a &  A^+ \\
A^- & t(a)
\end{array} \right]
\label{gdm1} \end{equation}
where $a$ is a $3\times 3$ complex matrix, $A^+,\  A^- \in \cc^3$, viewed as column and row vectors respectively and $t(a)$ denotes the trace of $a$.\\
The product of two such matrices is defined by:
\begin{equation}
\begin{array}{c}
\left [ \begin{array}{cc} a & A^+ \\ A^- & t(a)
\end{array}\right]
\left [ \begin{array}{cc} b & B^+ \\ B^- & t(b)
\end{array}\right] \\ =
\left [\begin{array}{cc} a b + A^+ \circ  B^- &
a B^+  + A^- \wedge B^- \\
A^- b + A^+ \wedge B^+ & t(a)t(b) + t(B^+ \circ A^-) \end{array} \right],
\label{alg}
\end{array}
\end{equation}
where
\be
A^+ \circ B^- = t(A^+ B^- )  I - t(I) A^+ B^-
\label{circ}
\ee
(with standard matrix products of row and column vectors and with $I$ denoting the $3\times 3$ identity matrix); $A
\wedge B$ is the standard vector product of $A$ and $B$. Notice that $t(X^+\circ Y^-) = 0$, hence we have an algebra.

In particular, we get a sub-algebra by imposing the $a$ matrices to be traceless. We use this algebra in order to
define the following adjoint representation $\rep$ of the Lie algebra $\gd$:
\begin{equation}
\left[
\begin{array}{cc}
a &  A^+ \\
A^- & 0
\end{array} \right]
\label{gdm2} \end{equation}
where $a \in \ad$, $A^+,\  A^- \in \cc^3$, viewed as column and row vector respectively.\\
Indeed, the commutator of two such matrices, using (\ref{alg}), can be computed to read :
\begin{equation}
\begin{array}{c}
\left [ \left[ \begin{array}{cc} a & A^+ \\ A^- & 0
\end{array}\right]
 ,
\left [ \begin{array}{cc} b & B^+ \\ B^- & 0
\end{array}\right]\right] \\ =
\left [\begin{array}{cc} [a, b] + A^+ \circ  B^- - B^+\circ A^- &
a B^+  - b A^+ + 2 A^- \wedge B^- \\
A^- b - B^-a + 2 A^+ \wedge B^+ & 0 \end{array} \right],
\label{comm}
\end{array}
\end{equation}

and therefore one is led to the following identifications of the $\gd$ generators shown above:
\be
\begin{array}{l}
\rep (d^\pm_k) = E_{k\pm 1 \  k\pm 2} \quad \text{(mod 3)} \ , \ k = 1,2,3\\
\rep(\sqrt{2} H_1) = E_{11} - E_{22} \qquad \rep(\sqrt{6} H_2) = E_{11} + E_{22} - 2 E_{33} \\
\rep(g^+_k) = E_{k 4} := \ekpe \quad \rep(g^-_k) = E_{4 k} := \ekme \quad  , \ k = 1,2,3
\end{array}
\ee
where $E_{ij}$ denotes the matrix with all zero elements except a $1$ in the $\{ij\}$ position: $(E_{ij})_{k\ell} = \delta_{ik}\delta_{j\ell}$ and $\ekpe$ are the standard basis vectors of $\cc^3$ ($\ekme$ denote their transpose).

On the other hand, a direct calculation shows that:
\be
\rep ([X,Y]) = [\rep(X),\rep(Y)] \quad X,Y \in \gd
\ee
thus proving that $\rep$ (which is obviously linear) is indeed a representation.

\begin{figure}
\begin{center}
\includegraphics{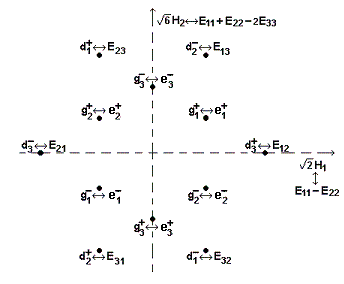}
\caption{Diagram of roots of $\gd$ with corresponding generators and {\it matrix-like} elements\label{fig:g2m}}
\end{center}
\end{figure}

It is useful to extend this correspondence to the roots of $\gd$, obtaining the pictorial view of the diagram in figure \ref{fig:g2m}.

For future use, we here explicitly associate a matrix in the form \eqref{gdm2} to the derivation $D_{c,d}$ for $c,d \in \oo$.
Let us define $e_{jk} := D_{\varepsilon^-_k , \varepsilon^+_j}$ (notice the switch of indices).
A straightforward calculation shows that, for $c,d \in \oo$, $c = \alpha^\pm_0 \rho^\pm + v^\pm_k \ekpm$, $d = \beta^\pm_0 \rho^\pm + w^\pm_k \ekpm$, it holds that:
\bea{rl}
D_{c,d} &= \frac13  \left( (\alpha^\pm_0  - \alpha^\mp_0) w^\pm_k  -  (\beta^\pm_0  - \beta^\mp_0) v^\pm_k - (v^\mp \wedge w^\mp)_k \right) g^\pm_k \\
&\phantom{=} + (v^-_k w^+_j - v^+_j w^-_k) e_{jk}
\label{eij}
\eea
Notice that $\rep(e_{ij}) = E_{ij}$ for $i\neq j=1,2,3$, whereas
$$\rep(e_{11}) = \rep(\frac{\sqrt2}2 H_1 + \frac{\sqrt6}6 H_2 = \frac13 (2 E_{11} - E_{22} - E_{33})$$
$$\rep(e_{22}) = \rep(- \frac{\sqrt2}2 H_1 + \frac{\sqrt6}6 H_2 = \frac13 (- E_{11} - E_{22} +2 E_{33})$$
$$\rep(e_{33}) = \rep(- \frac{\sqrt6}3 H_2 = \frac13 (-E_{11} - E_{22} + 2 E_{33})$$
We thus obtain from \eqref{eij}:
\bea{ll}
\rep(D_{c,d}) &=
\left( \begin{array}{cc}
D_{11} & D_{12} \\ D_{21} & 0 \end{array} \right), \qquad \text{where}\\
D_{11} &= -\frac13  (v^-_i w^+_i - v^+_i w^-_i) I +  (v^-_j w^+_i - v^+_i w^-_j) E_{ij}  \\
D_{12} &=  \frac13  \left( (\alpha^+_0  - \alpha^-_0) w^+  -  (\beta^+_0  - \beta^-_0) v^+ - (v^- \wedge w^-) \right) \\
D_{21} &=  \frac13  \left( (\alpha^-_0  - \alpha^+_0) w^-  -  (\beta^-_0  - \beta^+_0) v^- - (v^+ \wedge w^+) \right)
\label{dcd}
\eea

We introduce the following action of $\rep(\gd)$ on the  octonions represented by Zorn matrices:
\be
\begin{array}{c}
\left[ \left [ \begin{array}{cc} a & A^+ \\ A^- & 0
\end{array}\right] ,
\left[
\begin{array}{cc}
\alpha_0^+ &  v^+ \\
v^- & \alpha_0^-
\end{array} \right] \right] \\ \\=
\left [\begin{array}{cc} -  v^-  A^+  +   A^- v^+&
a v^+  + (\alpha^-_0 - \alpha^+_0) A^+ - A^- \wedge v^- \\
- v^- a  - (\alpha^-_0 - \alpha^+_0) A^- -  A^+ \wedge v^+ &   v^- A^+ - A^-   v^+  \end{array} \right]
\label{g21a}
\end{array}
\ee

We see that $\rep(\gd)$ acts non-trivially on traceless octonions, hence we can write $\alpha_0^+ = - \alpha_0^-$ to get a 'matrix-like' expression of the 7-dimensional (fundamental) representation $\mathbf{7}$ of $\gd$.

A direct calculation confirms that the action (\ref{g21a}) corresponds to the action of the $\gd$ generators on the octonions shown above.

It can also be shown that the action (\ref{g21a}) is indeed a derivation of the octonions, confirming that $Der(\oo)=\gd$. The only ingredients needed for the proof are identities from elementary 3-dimensional geometry, like
$(A\wedge B)\cdot C = (C\wedge A)\cdot B$ or $(A\wedge B) \wedge C = (A\cdot B) C - (B\cdot C) A$, plus the following identity for a $3\times 3$ traceless matrix $a$:
$$\sum_j a_{ij} \epsilon_{jk\ell} + a_{kj} \epsilon_{ij\ell} + a_{\ell j} \epsilon_{ikj} = 0$$.

\section{$\mathbf{n}=1$ : Matrix representation of $\fq$}\label{sec:f4}

We introduce in this section the representation $\rep$ of $\fq$ in the form of a matrix. For $\fqe \in \fq$:

\begin{equation}
\rep(\fqe) = \left ( \begin{array}{cc} a\otimes I + I \otimes \au & \vxp \\ \vxm & -I\otimes \aut
\end{array}\right),
\label{mfq}
\end{equation}
where $a\in \adu$, $\au \in \add$ (the superscripts being merely used to distinguish the two copies of $\ad$) $\aut$ is the transpose of $\au$,  $I$ is the $3\times 3$ identity matrix, $\vxp \in \cc^3 \otimes \jotu$,  $\vxm \in \cc^3 \otimes \jobtu$ :
$$ \vxp := \left( \begin{array}{c} x_1^+ \\ x_2^+ \\ x_3^+ \end{array} \right) \quad  \vxm := (x_1^- , x_2^- , x_3^- )\ ,  \ x_i^+ \in \jotu \ \ x_i^- \in \jobtu \ , \quad i=1,2,3$$

The commutator is set to be:

\be
\begin{array}{c}
\left[
\left ( \begin{array}{cc} a\otimes I + I \otimes \au & \vxp \\ \vxm & -I\otimes \aut
\end{array}\right) ,
\left ( \begin{array}{cc} b\otimes I + I \otimes \bu &\vyp \\ \vym & -I\otimes \but
\end{array}\right) \right] \\  \\ :=
\left (\begin{array}{cc} C_{11} & C_{12}\\
C_{21} & C_{22}
 \end{array} \right) \hfill
\label{fqcom}
\end{array}
\ee

where:

\be
\begin{array}{ll}
C_{11} &= [a,b] \otimes I + I \otimes [\au,\bu] + \vxp \diamond \vym - \vyp \diamond \vxm \\ \\
C_{12} &=  (a \otimes I) \vyp -  (b \otimes I) \vxp + (I \otimes \au) \vyp + \vyp (I \otimes \aut) \\
 &\phantom{:=} - (I \otimes \bu) \vxp - \vxp (I \otimes \but) +  \vxm \times \vym \\ \\
C_{21} &= - \vym (a \otimes I)  +  \vxm (b \otimes I) - (I \otimes \aut) \vym - \vym (I \otimes \au)  \\
&\phantom{:=} + (I \otimes \but) \vxm + \vxm (I \otimes \bu) +  \vxp \times \vyp \\ \\
C_{22} &=  I \otimes [\aut,\but] + \vxm \bullet \vyp - \vym \bullet \vxp
\end{array}
 \label{comrel}
\ee
with the following definitions :
\be
\begin{array}{ll}
\vxp \diamond \vym &:= \left(\frac13 t(x^+_i, y^-_i) I - t(x^+_i,y^-_j) E_{ij}\right) \otimes I +\\
&\phantom{:=} I \otimes \left(\frac13 t(x^+_i, y^-_i ) I - x^+_i y^-_i \right) \\ \\
\vxm \bullet \vyp &:= I \otimes (\frac13 t(x^-_i,y^+_i) I -  x^-_i y^+_i) \\  \\
(\vxpm \times \vypm)_i &:= \epsilon_{ijk}[x_j^\pm y_k^\pm + y_k^\pm x_j^\pm -x_j^\pm t(y_k^\pm) - y_k^\pm t(x_j^\pm) \\
&\phantom{:=}- (t(x_j^\pm, y_k^\pm) - t(x_j^\pm) t( y_k^\pm)) I]  \\
&:= \epsilon_{ijk} (x_j^\pm \# y_k^\pm)
\end{array}
 \label{not1}
\ee
Notice that:
\begin{enumerate}
\item $x \in \jotu$ is a symmetric complex matrix;
\item writing $\vxp \diamond \vym := c \otimes I + I\otimes \cu$ we have that both $c$ and $\cu$ are traceless hence $c, \cu \in \ad$, and indeed they have $8$ (complex) parameters, and $\vym \bullet \vxp = I\otimes \cut$;
\item terms like $(I \otimes \au) \vyp + \vyp (I \otimes \aut)$ are in $\cc^3\otimes \jotu$, namely they are matrix valued vectors with symmetric matrix elements;
\item the {\it sharp} product $\#$ of $\jotu$ matrices appearing in $\vxpm \times \vypm$ is the fundamental product in the theory of Jordan Algebras, introduced in section \ref{sec:jp}.
\end{enumerate}
In order to prove that $\rep$ is a representation of the Lie algebra $\fq$ we make a comparison with Tits' construction of the fourth row of the magic square, \cite{tits2} \cite{freu1}.
If $\jo_0$ denotes the traceless elements of $\jo$, $\oo_0$ the traceless octonions (the trace being defined by $t(a) := a + \bar a \ \in \cc$, for $a\in \oo$
where the bar denotes the octonion conjugation - that does not affect the field $\cc$ - ), it holds that :
\be
\fq = Der(\oo) \oplus (\oo_0 \otimes \jo_0) \oplus Der(\jo)
\ee
with commutation rules, for $D\in Der(\oo) = \gd $, $c,d \in \oo_0 $, $x,y \in \jo_0$, $E\in Der(\jo)$, given by:
\be
\begin{array}{l}
\left[ Der(\oo ) , Der(\oo) \right]  = Der(\oo)  \\ \\
\left[ Der(\jo) , Der(\jo)\right]  = Der(\jo)  \\ \\
\left[ Der(\oo), Der(\jo)\right]  = 0  \\ \\
\left[ D, c\otimes x \right] = D(c) \otimes x  \\ \\
\left[ E, c\otimes x \right] = c \otimes E(x)  \\ \\
\left[ c\otimes x, d \otimes y\right] = t(x y) D_{c,d} + 2 (c \ast d)  \otimes (x\ast y)  + \frac12 t(c d)  [x,y]
\end{array}
\label{tc}
\ee

\noindent
where , $\oo_0\ni c\ast d = c d - \frac12 t(c d)$, $\jo_0\ni x\ast y = \frac12 (x y + y x) - \frac13 t(x y) I$, $\jo = \jotu$.

The derivations of $\jo$ are inner: $Der(\jo) = [L(\jo),L(\jo)]$ where $L$ stands for the left (or right) multiplication with respect to the Jordan product: $L_x y = \frac12 (x y + y x)$.
In the case under consideration, the product $x,y \to xy$ is associative and $[L_x,L_y] z = \frac14 [[x,y],z]$. Since $[x,y]$ is antisymmetric, then $Der(\jo) = so(3)_\cc \equiv \aun$.

We can thus put forward the following correspondence:
\be
\begin{array}{ll}
\rep(D) &= \left(\begin{array}{cc} a \otimes I  &  \frac13 tr(\vxp) \otimes I \\ \frac13 tr(\vxm) \otimes I & 0 \end{array}\right) \\ \\
\rep(E)  &= \left(\begin{array}{cc} I \otimes a^A_1  &  0 \\ 0 & I \otimes a^A_1  \end{array}\right)  \\ \\
\rep(\ekp \otimes \jo_0) &= \left(\begin{array}{cc} 0  &  \vxp_k - \frac13 tr(\vxp_k) \otimes I  \\ 0 & 0   \end{array} \right) \\ \\
\rep(\ekm \otimes \jo_0) &= \left(\begin{array}{cc} 0  &  0 \\ \vxm_k - \frac13 tr(\vxm_k) \otimes I  & 0   \end{array} \right) \\ \\
\rep((\rho^+ - \rho^-) \otimes \jo_0) &= \left(\begin{array}{cc} I \otimes a^S_1  &  0 \\ 0 & - I \otimes a^S_1 \end{array} \right),
\end{array}
\label{corr}
\ee
where $a^A_1$ and $a^S_1$ are the antisymmetric and symmetric parts of $a_1$, and $$tr(\vxp) := \left( \begin{array}{c} t(x^+_1) \\  t(x^+_2) \\  t(x^+_3) \end{array} \right) \ , \quad  tr(\vxm) = ( t(x^-_1) ,  t(x^-_2) ,  t(x^-_3))$$
with $\vxpm_k$ denoting a matrix-valued vector whose $k$-th component is the only non-vanishing one.

\subsection{Comparison with Tits' construction}\label{subsec:Tits-comp}

It is here worth commenting that there is some apparent difference between the way we write Tits' construction, \eqref{tc}, and the way it is written in the mathematical literature; see for instance \cite{jacob2}, page93.

Firstly, we have the operators acting from the left, contrary to the action from the right often used by mathematicians. This implies that the third and fourth commutators in (\ref{tc}) are written in the reverse order.
Moreover, the last commutator of (\ref{tc}) is instead written in \cite{jacob2} (using the superscript ${}^\top$ in order to distinguish it from ours) as follows :
\be
 {\left[ c\otimes x, d \otimes y\right]}^\top = \frac1{12} t(x y) D^\top_{c,d} + ((c \ast d)  \otimes (x\ast y))^\top  + \frac12 t(c d)  [L_x,L_y].
\label{jaco}
\ee
Furthermore, we observe that we have defined the derivation $D_{a,b}= \frac13D^\top_{a,b}$  (up to a sign due to left \textit{versus} right action).
Because of this and the fact that $[L_x,L_y] z = \frac14 [[x,y],z]$, we have 4 times the first and third terms in \eqref{jaco}, and 2 times the middle one.
However, these factors can be reabsorbed by changing $\rep(c\times x) \to \frac12 \rep(c\times x)$, thus proving the equivalence of the two ways of writing all the commutation relations.

\subsection{$\rep$ as a representation of $\fq$}\label{subsec:proof-f4}

By exploiting the correspondence \eqref{corr}, we now prove in six steps that the commutators \eqref{comrel} satisfy \eqref{tc}, thus proving the following

{\bf{Theorem}} : $\rep$ realizes the adjoint representation of $\fq$.
\\
{\bf{Proof}} :
\noindent
1) $\left[ Der(\oo ) , Der(\oo) \right]  = Der(\oo) $\\

In order to prove this first step, let us denote by $A^\pm$ and $B^\pm$ the $\cc^3$ vectors $\frac13 tr(\vxpm)$ and $\frac13 tr(\vypm)$ respectively. Then, we have to compute:\\

\be
\begin{array}{c}
\left[
\left ( \begin{array}{cc} a\otimes I  & A^+ \otimes I \\ A^- \otimes I & 0
\end{array}\right) ,
\left ( \begin{array}{cc} b\otimes I  & B^+ \otimes I \\ B^- \otimes I & 0
\end{array}\right)\right] :=
\left (\begin{array}{cc} C_{11} & C_{12}\\
C_{21} & C_{22}
 \end{array} \right) \hfill
\end{array}
\label{comf4}
\ee

Let us calculate some terms separately:
\bes
A^+\otimes I \diamond B^-\otimes I = (A^+_iB^-_i I - 3 A^+_i B^-_j E_{ij}) \otimes I + I \otimes ( \frac13 A^+_i B^-_i t(I) I - A^+_i B^-_i I );\\
\ees
the first bracket on the right-hand side is $A^+\circ B^-$, as defined in \eqref{circ}, whereas the second one vanishes.

\bes
(A^+ \otimes I) \times (B^+ \otimes I)_i = \epsilon_{ijk} A^+_j B^+_k (2 I  - 2 t(I) I - (t(I) - t(I)^2)I) = 2 (A^+\wedge B^+)_i  I;
\ees
similarly with $A^-$ and $B^-$, hence
\bes
(A^\pm \otimes I) \times (B^\pm \otimes I) = 2 (A^\pm \wedge B^\pm) \otimes I.
\ees
Therefore, we obtain
\be
\begin{array}{ll}
C_{11} &= ([a,b] + A^+ \circ B^- - B^+ \circ A^-) \otimes I\\
C_{12} &= ( a   B^+  -  b A^+  + 2 A^- \wedge B^-) \otimes I \\
C_{21} &= ( - B^-  a  + A^- b  + 2 {A^+} \wedge {B^+}) \otimes I   \\
C_{22} &=  0,
\end{array}
\ee
which make the commutation relations \eqref{comf4} correspond to those of $\gd$ introduced in \eqref{comm}.\\

\noindent
2), 3) $\left[ Der(\jo) , Der(\jo)\right]  = Der(\jo)$ , $\left[ Der(\oo), Der(\jo)\right]  = 0$  \\

One can prove this in a straightforward way, \textit{e.g.} by explicit computation.\\

\noindent
4) $\left[ D, c\otimes x \right] = D(c) \otimes x $ \\

In order to prove this, let us write $c = \alpha (\rho^+ - \rho^-) + v_k^\pm  \ekpm \in \oo_0$ (summed over $\pm$ and $k$), and let us consider $V^\pm \in \cc^3$ with components $v^\pm_k$. Then, we have to calculate:
\be
\begin{array}{c}
\left[
\left ( \begin{array}{cc} a\otimes I  & A^+ \otimes I \\ A^- \otimes I & 0
\end{array}\right) ,
\left ( \begin{array}{cc} \alpha I \otimes x  & V^+ \otimes x \\ V^- \otimes x & -\alpha I \otimes x
\end{array}\right)\right]  :=
\left (\begin{array}{cc} C_{11} & C_{12}\\
C_{21} & C_{22}
 \end{array} \right) \hfill
\end{array}
\ee
Let us calculate some terms separately:
\bes
\begin{aligned}
A^+\otimes I \diamond V^-\otimes x &= (\frac13 A^+_i v^-_i t(x) I -  A^+_i v^-_j t(x)E_{ij}) \otimes I \\
&\phantom{=}+ I \otimes (\frac13 A^+_i v^-_i t(x) I - A^+_i v^-_i x )\\
 &= - I \otimes  A^+_i v^-_i x,
\end{aligned}
\ees
since $t(x) = 0$ by hypothesis.
\bes
(A^+ \otimes I) \times (V^+ \otimes I)_i = \epsilon_{ijk} A^+_j v^+_k (2 x  - x t(I) ) = -(A^+\wedge v^+)_i  x,
\ees
once again because $t(x)=0$. Similarly with $A^-$ and $v^-$, hence
\bes
(A^\pm \otimes I) \times (V^\pm \otimes x) = - (A^\pm \wedge V^\pm) \otimes x
\ees
Consequently, we obtain
\be
\begin{array}{ll}
C_{11} &= I \otimes (- v^-_i A^+_i + A^-_i v^+_i) x \\
C_{12} &= ( a   V^+ - 2 \alpha A^+  -  A^- \wedge B^-) \otimes x \\
C_{21} &= ( - V^-  a  + 2 \alpha A^- - {A^+} \wedge {V^+}) \otimes x   \\
C_{22} &=  I \otimes (v^-_i A^+_i - A^-_i v^+_i) x,
\end{array}
\ee
which is the $\gd$ action on $c\in \oo_0$  introduced in \eqref{g21a} tensored with $x$.\\

\noindent
5) $\left[ E, a\otimes x\right] = a \otimes E(x)$ (this can be proved by explicit computation).\\

\noindent
6) $\left[ c\otimes x, d \otimes y\right] =  t(x y) D_{c,d} + 2 (c\ast  d) \otimes (x\ast y)  + \frac12 t(c d)[x,y] $ \\\label{sei}

Let us use notations analogous to the ones in the proof of 4). Then, we have to compute:
\be
\begin{array}{c}
\left[
\left ( \begin{array}{cc} \alpha I \otimes x  & V^+ \otimes x \\ V^- \otimes x & -\alpha I \otimes x
\end{array}\right) ,
\left ( \begin{array}{cc} \beta I \otimes y  & W^+ \otimes y \\ W^- \otimes y & -\beta I \otimes y
\end{array}\right)\right] :=
\left (\begin{array}{cc} C_{11} & C_{12}\\
C_{21} & C_{22}
 \end{array} \right) \hfill
\end{array}
\ee

Let us calculate some terms separately:
\bes
\begin{array}{ll}
(V^+\otimes x) \diamond (W^-\otimes y) &= (\frac13 v^+_i w^-_i t(xy) I -  v^+_i w^-_j t(xy)E_{ij}) \otimes I \\
	&\phantom{=} + I \otimes (\frac13 v^+_i w^-_i t(xy) I - v^+_i w^-_i xy )
\end{array}
\ees

\bes
(V^+ \otimes x) \times (W^+ \otimes y)_i = \epsilon_{ijk} v^+_j w^+_k (x y + y x - t(x y) I) )
\ees
being $t(x)=t(y)=0$. Similarly with $A^-$ and $v^-$, hence
\bes
(V^\pm \otimes x) \times (W^\pm \otimes y) = - (V^\pm \wedge W^\pm) \otimes (x y + y x - t(x y) I)
\ees
Therefore, one obtains
\bes
\begin{array}{ll}
C_{11} &= \left( \frac13 (v^+_i w^-_i  -  w^+_i v^-_i) t(xy) I - (v^+_i w^-_j - w_i^+ v_j^-) t(xy)E_{ij} \right)\otimes I \\
      &\ +\  I \otimes \left( \alpha \beta [x,y]  + \frac13 (v^+_i w^-_i  -  w^+_i v^-_i ) t(xy) I - v^+_i w^-_i  x y +w^+_i v^-_i  y x \right) \\ \\
 &= \left( \frac13 (v^+_i w^-_i  -  w^+_i v^-_i) t(xy) I - (v^+_i w^-_j - w_i^+ v_j^-) t(xy)E_{ij} \right)\otimes I \\
      &\ +\  I \otimes \left( \left(\alpha \beta - \frac12 (v^-_i w^+_i + v^+_i w^-_i)\right)  [x,y]  \right. \\
	&\ \left. + (v^-_i w^+_i  -  w^-_i v^+_i )( \frac12(x y + y x) - \frac13 t(xy) I ) \right) \\ \\
C_{12} &= ( \alpha   W^+ - \beta V^+) \otimes (x y + yx) + (V^- \wedge W^-) \otimes (x y + y x - t(x y) I) \\
	  &= 2( \alpha   W^+ - \beta V^+ + V^- \wedge W^-) \otimes (\frac12(x y + y x) - \frac13 t(xy) I) \\
	  &\phantom{=} + (  2 \alpha   W^+ - 2 \beta V^+-V^- \wedge W^-) \otimes \frac13 t(x y) I)
\end{array}
\ees
with similar results for $C_{21}$ and $C_{22}$.

Finally, for $c,d \in \oo_0$, $c = \alpha (\rho^\pm - \rho^-) + v^\pm_k \ekpm$, $d = \beta (\rho^+-\rho^-) + w^\pm_k \ekpm$ (summed over $\pm$ and $k$),
$V^\pm, W^\pm \in \cc^3$ with components $v^\pm_k, v^\pm_k$, it can be computed that :
\bea{ll}
c\ast d &= \frac12 (v^-_k w^+_k - v^+_k w^-_k)(\rho^+ - \rho^-) + (\pm\alpha w^\pm_k \mp \beta v^\pm_k + (V^\mp\wedge W^\mp)_k)\ekpm\\
\frac12 t(cd) &= \alpha \beta -\frac12 (v^-_k w^+_k - v^+_k w^-_k)
\label{tast}
\eea

From \eqref{tast} and \eqref{dcd} we obtain indeed the proof of 6).\\

This completes the proof that $\rep$ \eqref{mfq} is the adjoint representation of $\fq$. $\blacksquare $\\

Notice that \eqref{mfq} reproduces the well known branching rule of the adjoint of $\fq$
with respect to its maximal and non-symmetric subalgebra $\adu\oplus\add$:
\begin{equation}
\mathbf{52}=\left( \mathbf{8},\mathbf{1}\right) +\left( \mathbf{1},\mathbf{8}%
\right) +\left( \mathbf{3}, \overline{\mathbf{6}}\right)+\left( \overline{\mathbf{3}}, \mathbf{6}\right).
\end{equation}

It is here worth anticipating that in section \ref{sec:jacobi} we prove the Jacobi identity in the more general case of $\est$, which includes in an obvious manner this case of $\fq$. The validity of the Jacobi identity, together with the fact that the representation $\rep$ fulfills the root diagram of $\fq$ (the proof is straightforward, and it can also be considered as a particular case of the proof given at the end of section \ref{sec:e6} for $\es$) proves in an alternative way that $\rep$ is indeed a representation of $\fq$.


\section{$\mathbf{n}=2$ : Matrix representation of $\es$}\label{sec:e6}

We present in this section the representation $\rep$ of $\es$ in the form of a matrix. We have to complexify the Jordan structure with respect to $\fq$. We introduce
the imaginary unit $\uu$ - leaving $\imi$ as the imaginary unit of the base field. In particular, $\jotd$ is Hermitian with respect to the $\uu$-conjugation, and we are going to denote such an Hermitian conjugation with the symbol $\dagger$ throughout.

In a similar fashion to \eqref{mfq}, for $\fqe \in \es$, we thus write:
\begin{equation}
\rep(\fqe) = \left ( \begin{array}{cc} a\otimes I + I \otimes \au & \vxp \\ \vxm & -I\otimes \aud
\end{array}\right)
\label{mes}
\end{equation}
where $a\in \adf$, $\au \in \adgu \oplus \uu \adgd$, $\aud$ is the Hermitian conjugate of $\au$ (with respect to $\uu$),  $I$ is the $3\times 3$ identity matrix, $\vxp \in \cc^3 \otimes \jotd$,  $\vxm \in \cc^3 \otimes \jobtd$ :
$$ \vxp = \left( \begin{array}{c} x_1^+ \\ x_2^+ \\ x_3^+ \end{array} \right) \quad  \vxm = (x_1^- , x_2^- , x_3^- )\ ,  \ x_i^+ \in \jotd \ \ x_i^- \in \jobtd \ , \quad i=1,2,3$$

The commutator of two such matrices is the same as for $\fq$, with $\dagger$ instead of $T$ ({\textit{cfr.}} \eqref{fqcom}):
\be
\begin{array}{c}
\left[
\left ( \begin{array}{cc} a\otimes I + I \otimes \au & \vxp \\ \vxm & -I\otimes \aud
\end{array}\right) ,
\left ( \begin{array}{cc} b\otimes I + I \otimes \bu &\vyp \\ \vym & -I\otimes \bud
\end{array}\right) \right] \\  \\ :=
\left (\begin{array}{cc} C_{11} & C_{12}\\
C_{21} & C_{22}
 \end{array} \right) \hfill
\label{escom}
\end{array}
\ee
where:
\be
\begin{array}{ll}
C_{11} &= [a,b] \otimes I + I \otimes [\au,\bu] + \vxp \diamond \vym - \vyp \diamond \vxm \\ \\
C_{12} &=  (a \otimes I) \vyp -  (b \otimes I) \vxp + (I \otimes \au) \vyp + \vyp (I \otimes \aud) \\
 &\phantom{:=} - (I \otimes \bu) \vxp - \vxp (I \otimes \bud) +  \vxm \times \vym \\ \\
C_{21} &= - \vym (a \otimes I)  +  \vxm (b \otimes I) - (I \otimes \aud) \vym - \vym (I \otimes \au)  \\
&\phantom{:=} + (I \otimes \bud) \vxm + \vxm (I \otimes \bu) +  \vxp \times \vyp \\ \\
C_{22} &=  I \otimes [\aud,\bud] + \vxm \bullet \vyp - \vym \bullet \vxp
\end{array}
 \label{comreles}
\ee
with products defined as in \eqref{not1}.

Notice that:
\begin{enumerate}
\item $x \in \jotd$ is a Hermitian matrix (with respect to $\uu$) over the complex field (with imaginary unit $\imi$);
\item by writing $\au \in \adgu \oplus \uu \adgd$ we state that $\au$ is the sum of a traceless skew-Hermitian matrix and a traceless hermitian matrix (namely a matrix in $\jo_0$, with $\jo = \jotd$), hence $\au \in \sltc$ is a generic $3\times 3$ traceless matrix over $\cc \otimes \cc$;
\item writing $\vxp \diamond \vym := c \otimes I + I\otimes \cu$ we have that both $c$ and $\cu$ are traceless, $c \in \ad$ and $\cu \in \sltc$, and $\vym \bullet \vxp = I\otimes \cud$; if $x,y \in \jotd$, then $\cc \ni t(x, y)  = t(xy)$, and $\cu$ has indeed 16 (complex) parameters.  It is here worth anticipating that this will not be the case for $\jotq$ and $\joto$, as we shall stress in the next sections on $\est$ and $\eo$;
\item terms like $(I \otimes \au) \vyp + \vyp (I \otimes \aud)$ are in $\cc^3\otimes \jotd$, namely they are matrix valued vectors with Hermitian matrix elements;
\item the correspondence between matrix elements in \eqref{mes} and Tits' construction is similar to the one shown in \eqref{corr} and is omitted here;
\item the Jacobi identity can be demonstrated as a particular case of the proof for $\est$, shown in section \ref{sec:jacobi}. The validity of the Jacobi identity, together with the fact that the representation $\rep$ fulfills the root diagram of $\es$, as we show next, prove that $\rep$ \eqref{mes} is the adjoint representation of $\es$. $\blacksquare $
\end{enumerate}

\medskip

As regards the counting of parameters, we refer to our comment in the introduction about the use of $\cc$ as base field.\\

We end this section with the correspondence between the roots of $\es$ and the matrix elements in \eqref{mes}.

The roots of $\es$ can be written in terms of an orthonormal basis $\{ k_i \ | \ i= 1, \dots , 6 \}$ as, \cite{pt1}:
\beas{llc}
\es & &\textbf{72 roots}\\
&\pm k_i \pm  k_j \qquad  i\ne j = 1, \dots ,5 &  4\times \binom{5}{2} = 40 \\
&\frac{1}{2} (\pm k_1 \pm k_2 \pm k_3 \pm k_4  \pm k_5  \pm \sqrt{3}  k_6)^*  & 2^5= 32\\
& ^*\text{ [odd number of + signs]}
\eeas

We refer to Figure \ref{fig:e6roots} and write the roots associated with the highest weight $\jotd$ as:

\begin{figure}
\begin{center}
\includegraphics{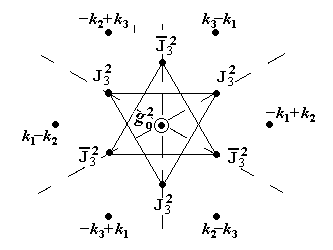}
\caption{Roots of $\es$ projected on the plane of $\gd$}\label{fig:e6roots}
\end{center}
\end{figure}

\bea{l}
- k_1 \pm k_4  \; , \qquad - k_1 \pm k_5 \; , \qquad k_2 + k_3\\
\frac{1}{2} ( - k_1 + k_2 + k_3 + k_4 - k_5 - \sqrt{3}  k_6)  \\ \frac{1}{2} ( - k_1 + k_2 + k_3 + k_4 + k_5 + \sqrt{3}  k_6)\\
\frac{1}{2} ( - k_1 + k_2 + k_3 - k_4 + k_5 - \sqrt{3}  k_6) \\ \frac{1}{2} ( - k_1 + k_2 + k_3 - k_4 - k_5 + \sqrt{3}  k_6)
\eea

The other $\jotd$'s correspond to a cyclic permutation of $k_1,k_2,k_3$, and each $\jobtd$ in a Jordan pair $(\jotd ,\jobtd)$ corresponds to the roots of a $\jotd$ with opposite signs.

The subalgebra $\gzd \simeq \ad \oplus \ad$ has roots:
\bea{l}
\pm (k_4 + k_5) \\ \pm \frac{1}{2} ( k_1 + k_2 + k_3 - k_4 - k_5 - \sqrt{3}  k_6)\\ \pm\frac{1}{2} ( k_1 + k_2 + k_3 + k_4 + k_5 - \sqrt{3}  k_6),
\eea
and
\bea{l}
\pm (k_4 - k_5) \\ \pm \frac{1}{2} ( k_1 + k_2 + k_3 - k_4 + k_5 + \sqrt{3}  k_6)\\ \pm \frac{1}{2} ( k_1 + k_2 + k_3 + k_4 - k_5 + \sqrt{3}  k_6).
\eea

Furthermore, the roots of $\adf$ relate in the standard way to the matrix elements of $a\otimes I$ in \eqref{mes}. The roots of each Jordan Pair project on the plane of $\adf$ according to Figure \ref{fig:e6roots}, as it can be easily checked. Therefore, we are only left with the roots corresponding to $\sltc$ and to each matrix element of a $\jotd$, say the highest weight one. The rest of the correspondence will readily follow.

The algebras $\adu$ and $\add$ are related to Tits' construction. Now, we twist them in the following way:
we denote by $\rho^\pm := \frac12 (1 \pm \imi\uu )$ and introduce $\adpm = \rho^\pm \sltc$.
Then, it follows that $(\rho^\pm)^2\ = \rho^\pm$ and $ \rho^\pm \rho^\mp = 0$. If $a\in \sltc$ then $a = a^+ + a^-$,
$a^\pm = \rho^\pm a$ and, if we write $a = a_r + \uu a_i$ (where $a_r$ and $a_i$ are the self-conjugate parts of $a$ with respect to $\uu$), one can easily check that $a^\pm = (a_r \mp \imi a_i) \rho^\pm$. Moreover, for $a,b\in \sltc$ then $[a,b] = [a^+ + a^-,b^++ b^-] = [a^+ ,b^+] + [a^-, b^-]$. Therefore $\ad^+$ and $\ad^-$ are both isomorphic to $\ad$ and $\sltc \simeq \ad^+ \oplus \ad^-$.

We now write $x\in \jotd = \alpha_i E_{ii} + a_{i, i+1} E_{i, i+1} + \bar a_{i, i+1} E_{i+1, i}$, where the indices run over $1,2,3$ mod(3), $\alpha\in \cc$ and $a_{ij}\in \cc \otimes \cc$. Obviously $\alpha_i = \alpha_i(\rho^+ +  \rho^-)$ and $a_{ij} =a_{ij}(\rho^+ +  \rho^-)$. The matrix $x$ is therefore in the linear span of the nine generators
\be
X_i = E_{ii} \ , \quad X^\pm_{i,i+1} := \rho^\pm E_{i, i+1} + \rho^\mp E_{i+1,i}
\ee

We fix the Cartan subalgebra of $\ad^+ \oplus \ad^-$ in the obvious way, by introducing the Cartan generators
\be
H^\pm_{1,2} :=  \rho^\pm H_{1,2} \ , \  H_1 = {\scriptstyle \frac{\sqrt{2}}2} \left( E_{11} - E_{22}\right) \ ,
\ H_2 = {\scriptstyle\frac{\sqrt 6}6} \left( E_{11} + E_{22} - 2 E_{33}\right)
\ee

We let $H^+_1,H^+_2$  correspond to the axes along the directions of the unit vectors ${\scriptstyle\sqd} (k_4 + k_5)$,
$-{\scriptstyle\sqs} (k_1 + k_2 + k_3 - \sqrt{3}  k_6)$ and $H^-_1,H^-_2$ to ${\scriptstyle\sqd}(k_4 - k_5)$, $-{\scriptstyle\sqs}(k_1 + k_2 + k_3 + \sqrt{3}  k_6)$ respectively.

Consequently, we are all set to establish the correspondence between the roots and the generators of the highest weight $\jotd$, by exploiting the commutation rule \eqref{escom}. This is shown  in Table \ref{table:rowe}.
\begin{table}[h]
\caption{Roots and $\ad^+ \oplus \ad^-$ weights of the highest weight $\jotd$}
\centering
\begin{tabular}{c c c c}
\hline\hline
Root & Generator & $\ad^+$ weights & $\ad^-$ weights \\ [0.2ex] 
\hline 
$- k_1 + k_4$ & $X_1$ & $\sqd , \sqs$  \\
$\frac{1}{2} ( - k_1 + k_2 + k_3 + k_4 - k_5 - \sqrt{3}  k_6)$  & $X^+_{3 1}$ & $0 , - \sqst$ & $\sqd , \sqs$ \\
$- k_1 - k_5$  & $X^-_{1 2}$ & $- \sqd , \sqs$  \\
\hline
$\frac{1}{2} ( - k_1 + k_2 + k_3 + k_4 + k_5 + \sqrt{3}  k_6)$ & $X^-_{3 1}$ & $\sqd , \sqs$  \\
$ k_2 + k_3$ & $X_3$ & $0 , - \sqst$ & $0 , - \sqst$ \\
$\frac{1}{2} ( - k_1 + k_2 + k_3 - k_4 - k_5 + \sqrt{3}  k_6)$ & $X^+_{23}$ & $-\sqd , \sqs$ \\
\hline
$- k_1 + k_5$ & $X^+_{1 2}$ & $\sqd , \sqs$  \\
$\frac{1}{2} ( - k_1 + k_2 + k_3 - k_4 + k_5 - \sqrt{3}  k_6)$ & $X^-_{2 3}$ & $0 , - \sqst$  & $- \sqd , \sqs$ \\
$- k_1 - k_4$ & $X_{2}$ & $- \sqd , \sqs$  \\ [0.4ex] 
\hline\hline 
\end{tabular}
\label{table:rowe} 
\end{table}

We thus reproduce the well known branching rule of the adjoint of $\es$ with respect to its maximal and non-symmetric subalgebra $\adf \oplus \ad^+ \oplus \ad^-$:
\begin{equation}
\mathbf{78}=\left( \mathbf{8},\mathbf{1,1}\right) +\left( \mathbf{1},\mathbf{%
8,1}\right) +\left( \mathbf{1},\mathbf{1,8}\right) +\left( \mathbf{3},%
\mathbf{3},\mathbf{3}\right) +\left( \overline{\mathbf{3}},\overline{\mathbf{%
3}}\mathbf{,}\overline{\mathbf{3}}\right) ,
\end{equation}
with the exact correspondence of each single root with a matrix elements of \eqref{mes}.

It is intriguing to remark the quantum information meaning of the maximal
non-symmetric embedding of $\adf \oplus \ad^+ \oplus \ad^-$ into $\es$ has been investigated in \cite{DF-e6},
within the context of the so-called \textquotedblleft black hole - qubit
correspondence" \cite{BH-qubit-Refs}.


\section{$\mathbf{n}=4$ : Matrix representation of $\est$}\label{sec:est}
In the present section, we briefly mention how the results of the previous sections can be extended to the case of $\est$. Nothing different really occurs, as of course the Jordan algebras involved are of the type $\jotq$, whose elements associate with respect to the standard product of matrices.

For $\fqe \in \est$, we write:
\begin{equation}
\rep(\fqe) = \left ( \begin{array}{cc} a\otimes I + I \otimes \au & \vx \\ \vz & -I\otimes \aud
\end{array}\right)
\label{mest}
\end{equation}
where $a\in \adf$, $\au \in \af$, $\aud$ is the Hermitian conjugate of $\au$ (with respect to the quaternion units),  $I$ is the $3\times 3$ identity matrix, $\vx \in \cc^3 \otimes \jotq$,  $\vz \in \cc^3 \otimes \jobtq$ :
$$ \vx = \left( \begin{array}{c} x_1 \\ x_2 \\ x_3 \end{array} \right) \quad  \vz = (z_1, z_2 , z_3 )\ ,  \ x_i \in \jotq \ \ z_i \in \jobtq \ , \quad i=1,2,3$$

The commutator of two such matrices is formally the same as for $\es$ ({\textit{cfr.}} \eqref{escom}):

A few remarks are in order :
\begin{enumerate}
\item since $\af \simeq \sltq$ ({\textit{cfr. e.g.}} \cite{Townsend-Kugo, baez, Rios}), then $\au \in \af$ can be written as the sum of a skew-Hermitian matrix and a traceless Hermitian matrix in $\jo_0$, with $\jo = \jotq$; it is worth noting that $\sltq$ has 35 parameters, only one less than $gl(3,\qq)$ since the trace that is taken away from $\mathbf{gl}(3,\qq)$ is in $\cc$, not in $\cc \otimes \qq$;
\item writing $\vxp \diamond \vym := c \otimes I + I\otimes \cu$, we have that both $c$ and $\cu$ are traceless, $c \in \ad$ and $\cu \in \sltq$ (and indeed this latter has $35$ complex parameters), and $\vym \bullet \vxp = I\otimes \cud$; according to the previous point, the trace that we take away with the term $I\otimes \frac13 t(x_i^\pm , y_i^\mp) I$ in \eqref{not1} is in $\cc$ and $t(x, y) \ne t(x y)$ in general, due to non-commutativity;
\item terms like $(I \otimes \au) \vyp + \vyp (I \otimes \aud)$ are in $\cc^3\otimes \jotq$, namely they are matrix-valued vectors with Hermitian matrix elements;
\item the correspondence between matrix elements in \eqref{mest} and Tits' construction is similar to the one shown in \eqref{corr} (and commented in Sec. 6), and it is omitted here;
\item the Jacobi identity is demonstrated in section \ref{sec:jacobi};
\item the adjoint action in $\est$ implicitly provides us with the action of $\es$ on the fundamental representations $\mathbf{27}$ and $\overline {\mathbf{27}}$, since $\est \simeq \es \oplus \cc \oplus (\joto , \jobto)$.
\end{enumerate}

This last point deserves to be commented a little further, since it allows us to write the action of $\est$ by means of matrices that associate with respect to the standard matrix product instead of non-associative matrices of $\joto$.
In a way, we are nothing but doubling the procedure already implemented for $\gd$ in Sec. 4, where we have realized the octonions within a Zorn-type matrix, which was the basic structure for building up our representations. Here, we have to branch $\joto$ into associative matrices, and still recover non-associativity through a non-standard matrix product.

As a first step, we consider the $\es$ subalgebra. We select an imaginary unit in $\qq$, say $\uu$, and restrict $\jotq$ to $\jotd$ accordingly. Then, we pick two $\ad$'s inside $\af$ by setting $\ad^\pm = \rho^\pm \sltc \subset \sltq$, and $\rho^\pm = \frac12(1 \pm \imi \uu)$. We thus get the following $\es$ subalgebra of matrices:
\begin{equation}
\left ( \begin{array}{cc} a\otimes I + I \otimes ( \rho^+ \au^+ \oplus \rho^-\au^-) & \vx \\ \vz & -I\otimes ( \rho^- {\au^+}^T \oplus \rho^+ {\au^-}^T)
\end{array}\right)
\label{mesest}
\end{equation}
where $\au^\pm \in \ad$ and the vectors $\vx ,\ \vz$ have components $x_i \in \jotd ,\  z_i \in \jobtd (i=1,2,3)$.

We now introduce the nilpotent elements $\varepsilon^\pm := \rpm \ud$, so that a generic quaternion can be written as $\qq\ni q = q_0^\pm \rho^\pm + q^\pm \varepsilon^\pm$. The Jordan pair $(\mathbf{27}, \overline{\mathbf{27}})$ reads then:
\begin{equation}
\left ( \begin{array}{cc} I \otimes ( \varepsilon^+ \eta^+ + \varepsilon^-\eta^-) & \varepsilon^+ \vzp +   \varepsilon^- \vzm \\
\varepsilon^+ \vxip +   \varepsilon^- \vxim & I\otimes (\varepsilon^+ {\eta^+}^T +  \varepsilon^- {\eta^-}^T)
\end{array}\right)
\label{mesvs}
\end{equation}
where $\eta^\pm \in \glt$ are complex $3\times 3$ matrices, and $\vzp,\vzm,\vxip,\vxim \in \bun$ are skew symmetric complex matrix-valued vectors.

As a convention, we associate the $\mathbf{27}$ with all the '$+$' signs in \eqref{mesvs}, and thus the $\overline{\mathbf{27}}$ with the '$-$' signs.

The only parameter left with respect to an element of $\est$ is the sum of the diagonal elements of type $\lambda \uu = \lambda (\rop\! - \!\rom)$, ($\lambda\in \cc$), which is associated to the generator $\cc$ in the decomposition of $\est$ (see point 6 above).

The action of $\es$ on its $\mathbf{27}$ is:
\be
\begin{array}{c}
\left[
\left ( \begin{array}{cc} a\otimes I + I \otimes \rpm \au^\pm & \vx \\ \vz & -I \otimes \rmp {\au^\pm}^T
\end{array}\right) ,
\left ( \begin{array}{cc} I \otimes \epp\eta  &\epp\vzz \\ \epp \vxi &  I \otimes \epp \eta^T
\end{array}\right) \right] \\  \\ :=
\left (\begin{array}{cc} C_{11} & C_{12}\\
C_{21} & C_{22}
 \end{array} \right) \hfill
\label{escomest}
\end{array}
\ee
where, for $x_i,z_i \in \jotd$, $x_i = x_{i{\scriptscriptstyle+}} \rop + x_{i{\scriptscriptstyle-}} \rom$, $z_i = z_{i{\scriptscriptstyle+}} \rop + z_{i{\scriptscriptstyle-}} \rom$:
\be
\begin{array}{ll}
C_{11} &= \epp \left(I \otimes (\au^+ \eta - \eta \au^-) - (x_{i{\scriptscriptstyle+}} \xi_i - \zeta_i z_{i{\scriptscriptstyle-}})\right) \\ \\
C_{12} &=  \epp\left((a \otimes I) \vzz  + (I \otimes \au^+) \vzz + \vzz (I \otimes {\au^+}^T) \right.\\
 &\phantom{:=}\left. + \vx_{\scriptscriptstyle+} (I \otimes \eta^T) - (I \otimes \eta)\vx_{\scriptscriptstyle_-}  +  \vz \times \vxi \right) \\ \\
C_{21} &= \epp \left(- \vxi (a \otimes I)  - (I \otimes {\au^+}^T) \vxi - \vxi (I \otimes \au^-)\right.  \\
&\phantom{:=} \left.+ (I \otimes \eta^T) \vz + \vz (I \otimes \eta) +  \vx \times \vzz \right)\\ \\
C_{22} &=  \epp \left(I \otimes (-{\au^-}^T \eta ^T+ \eta^T {\au^+}^T) - (z_{i{\scriptscriptstyle+}} \zeta_i - \xi_i x_{i{\scriptscriptstyle-}})\right).
\end{array}
 \label{comrelesest}
\ee
Notice that if $x\in \jotd$, $x = x_{\scriptscriptstyle+} \rop + x_{\scriptscriptstyle-} \rom$, then $x = x^\dagger = x_{\scriptscriptstyle+}^T \rom + x_{\scriptscriptstyle-}^T \rop$ shows that $x_{\scriptscriptstyle+}^T = x_{\scriptscriptstyle-}$. Therefore
$x \cdot (\epp\zeta) = (x_{\scriptscriptstyle+} \zeta + \zeta x_{\scriptscriptstyle-})\epp$ where $(x_{\scriptscriptstyle+} \zeta + \zeta x_{\scriptscriptstyle-})$ is skew-symmetric. In particular, $t(x , \zeta) = 0$. It also holds that $(x_{i{\scriptscriptstyle+}} \xi_i - \zeta_i z_{i{\scriptscriptstyle-}})^T = (z_{i{\scriptscriptstyle+}} \zeta_i - \xi_i x_{i{\scriptscriptstyle-}})$, thus showing that $C_{22} = - C_{11}^\dagger$.
It can also be shown that $C_{12}$ and $C_{21}$ are the product of $\epp$ with a skew-symmetric complex matrix.

Analogous calculation can be performed for the $\overline{\mathbf{27}}$.

The action of the $\cc$ generator $\lambda (\rop - \rom)$ on the $\mathbf{27}$ and on the $\overline{\mathbf{27}}$ is just a multiplication by $2\lambda$ on the $\mathbf{27}$ and by $-2\lambda$ on the $\overline{\mathbf{27}}$.

We thus reproduce the well known branching rule of the adjoint of $\est$ with respect to its maximal and non-symmetric subalgebra $\ad \oplus \af$:
\begin{equation}
\mathbf{133}=\left( \mathbf{8},\mathbf{1}\right) +\left( \mathbf{1},\mathbf{%
35}\right) +\left( \mathbf{3},\overline{\mathbf{15}}\right) +\left(
\overline{\mathbf{3}},\mathbf{15}\right) .
\end{equation}


\section{Jacobi identity for $\fq , \es , \est$}\label{sec:jacobi}
An equivalent way of proving that $\rep$ (given by \eqref{mfq},\eqref{mes},\eqref{mest}) is a representation of $\fq , \es , \est$ respectively, is to directly prove the Jacobi identity for $\rho$, and check that one gets the root diagram of the corresponding Lie algebra.

We consider the most general setting of $\est$, which involves the Jordan algebra $\jotq$, with non-commutative, but associative matrix elements. The $\rep(\fq)$ and $\rep(\es)$ cases are obviously included as particular instances.

Recalling \eqref{mest}, we thus write:
\begin{equation}
\rep(\fqe_1) = \left ( \begin{array}{cc} a\otimes I + I \otimes \au & A^+ \\ A^- & -I\otimes \aud
\end{array}\right)
\label{mfqq}
\end{equation}
where $a\in \adf , \au \in\af \simeq \sltq$ and $A^+ , A^-$ are three-vectors with elements in $\jotq , \jobtq$. Similarly, one can define $\rep(\fqe_2)$ and $\rep(\fqe_3)$, by respectively replacing $a \to b$ and $a\to c$ in \eqref{mfqq}, and:
\be
[[\rep(\fqe_1),\rep(\fqe_2)],\rep(\fqe_3)]]+\text{cyclic permutations}
:= \left (\begin{array}{cc} \jac_{11} & \jac_{12}\\
\jac_{21} & \jac_{22}
 \end{array} \right)
\ee

In order for the Jacobi identity to hold for the matrix realization \eqref{mfqq} of the adjoint of $\est$, we have to prove that $\jac_{11} =\jac_{12} =\jac_{21}=\jac_{22} =0$.

After some algebra, one computes :
\beas{l}
 \jac_{11} =\\
\  [[a,b],c] \otimes I + I \otimes [[\au,\bu],\cu] + (A^+ \diamond B^- - B^+ \diamond A^-)(c\otimes I + I\otimes \cu) \\
 - (c\otimes I + I\otimes \cu)(A^+ \diamond B^- - B^+ \diamond A^-) \\
+ \left( (a \otimes I) B^+ -  (b \otimes I) A^+ + (I \otimes \au) B^+ + B^+ (I \otimes \aud) \right.\\
\left. - (I \otimes \bu) A^+ - A^+ (I \otimes \bud) +  A^- \times B^- \right) \diamond C^- \\
- C^+ \diamond \left(   - B^- (a \otimes I)  +  A^- (b \otimes I) - (I \otimes \aud) B^- - B^- (I \otimes \au) \right. \\
\left. + (I \otimes \bud) A^- + A^- (I \otimes \bu) +  A^+ \times B^+ \right) + \text{cyclic permutations}
\label{jazzz}
\eeas

The first two terms of \eqref{jazzz} vanish upon cyclic permutations because of the Jacobi identity in $\ad$ and $\af$. Let us consider then the terms in the r.h.s. of \eqref{jazzz} containing  $A^+ , B^- , c$; by denoting by $a_k , b_k \in \jotq$, ($k  = 1,2,3$) the components of $A^+$ and $B^-$, respectively, one computes that:
\beas{l}
[A^+ \diamond B^-,c\otimes I]  + ((c\otimes I) A^+ )\diamond B^- - A^+ \diamond (B^- (c\otimes I))\\
= [(\frac13 t(a_i,  b_i) I - t(a_i, b_j) E_{ij}) \otimes I) , c\otimes I] +
 [I \otimes (\frac13 t(a_i, b_i) I - a_i b_i), c\otimes I] \\
\pu + (\frac13 t(c_{ik}a_k,  b_i) I - t(c_{ik}a_k, b_j) E_{ij}) \otimes I) + I \otimes (\frac13 t(c_{ik}a_k, b_i) I - c_{ik}a_k b_i)\\
\pu -  (\frac13 t(a_i,  b_k c_{ki}) I - t(a_i, b_k c_{kj}) E_{ij}) \otimes I) + I \otimes (\frac13 t(a_i, b_kc_{ki}) I - a_i b_kc_{ki}) \\
= (- t(a_i, b_j) E_{ij} c + t(a_i, b_j) c E_{ij}- t(c_{ik}a_k, b_j) E_{ij} + t(a_i, b_k c_{kj}) E_{ij}) \otimes I\\
=  (- t(a_i ,b_k)  c_{kj} + t(a_k, b_j) c_{ik}- t(a_k, b_j) c_{ik} + t(a_i, b_k )c_{kj}) E_{ij} \otimes I \\
= 0.
\label{apbmc1}
\eeas

Next, we consider the terms in the r.h.s. of \eqref{jazzz} containing  $A^+ , B^- , c_1$.
They read:
\beas{l}
[A^+ \diamond B^-,I\otimes \cu]  + ((I\otimes \cu) A^+  + A^+ (I\otimes \cud))\diamond B^- \\
- A^+ \diamond ((I\otimes \cud) B^-  + B^-(I\otimes \cu))\\
= I\otimes (\cu a_i b_i - a_i b_i \cu) + \left( \frac13 t(\cu a_i + a_i \cud , b_i) I - t(\cu a_i + a_i \cud , b_j ) E_{ij}\right) \otimes I \\
\pu + I\otimes \left( \frac13 t(\cu a_i + a_i \cud , b_i) I -(\cu a_i + a_i \cud )b_i \right) \\
\pu -  \left( \frac13 t(a_i , \cud b_i + b_i \cu) I -t(a_i , \cud b_j + b_j \cu) E_{ij} \right) \otimes I \\
\pu - I \otimes \left( \frac13 t(a_i , \cud b_i + b_i \cu) I - a_i(\cud b_i + b_i \cu) \right)\\
=  I\otimes \left[ \frac23 \left( t(\cu a_i + a_i \cud , b_i) - t(a_i , \cud b_i + b_i \cu) \right) I \right. \\
\pu\left.- \left( t(\cu a_i + a_i \cud , b_j ) - t(a_i , \cud b_j + b_j \cu) \right) E_{ij} \right]
\label{apbmc2}
\eeas
In order to prove that the r.h.s. of \eqref{apbmc2} is zero, we write $\sltq \ni \cu = h + s$, where $h\in \jotq$ is Hermitian, and $s$ skew-Hermitian (with respect to quaternion conjugation). Note that the action $x \to sx+xs^\dagger = sx - xs$ is a derivation in $\jotq$. Therefore, by exploiting the identities \cite{McCrimmon,jacob2}:
\bea{l}
t(x,y\cdot z) = t(z,x\cdot y) \\
t(D x, y) + t(x, D y) = 0 \quad \text{where $D$ is a derivation in $\jotq$}
\label{idmj}
\eea
one proves that the terms under consideration in the r.h.s. of \eqref{apbmc2} sum up to zero.\\

Finally, we consider terms in the r.h.s. of \eqref{jazzz} which contain structures like $(A^- \times B^-) \diamond C^-$; they read:
\bea{l}
(A^- \times B^-) \diamond C^- + (B^- \times C^-) \diamond A^- + (C^- \times A^-) \diamond B^-\\
= \epsilon_{i \ell k} \left( \frac13 t (a_\ell \# b_k , c_i) I - t (a_\ell \# b_k , c_j)E_{ij} \right) \otimes I \\
\pu + I\otimes \epsilon_{i \ell k} \left( \frac13 t (a_\ell \# b_k , c_i) I -  (a_\ell \# b_k) c_i \right) + \text{cyclic permutations}\\
:= \Mu \otimes I + I \otimes \Md.
\label{apbmc3}
\eea

In order to show that $\Mu = \Md = 0$, we observe, after \cite{McCrimmon}, that $t(a\# b , c)$ is symmetric in $a,b,c$. Let us consider $\Mu$ first. For $i \ne j$, then either $j = \ell$ or $j=k$. The coefficient of $E_{ij}$ is therefore :
\beas{l}
\epsilon_{i j k} \left(t (a_j \# b_k , c_j) - t (a_k \# b_j , c_j) + t(b_j \# c_k , a_j) - t (b_k \# c_j , a_j) \right. \\
\pu \left.+  t(c_j \# a_k , b_j) - t (c_k \# a_j , b_j) \right) = 0
\eeas
\noindent For $i=j$, by summing over $i, \ell , k$ and using the notation $\tau_{\ell k i} :=  t (a_\ell \# b_k , c_i) +  t (b_\ell \# c_k , a_i) +  t (c_\ell \# a_k , b_i)$, one can easily check that:
$$\epsilon_{1 \ell k} \tau_{\ell k 1} =  \epsilon_{2 \ell k} \tau_{\ell k 2} = \epsilon_{3 \ell k} \tau_{\ell k 3} := \omega$$.
Thus :
\beas{l}
\epsilon_{i \ell k} \tau_{\ell k i}(\frac13 I - E_{ii}) = \omega (I - E_{11} - E_{22} - E_{33}) = 0
\eeas

This proves that $\Mu = 0$. For what concerns $\Md$, we observe that $\frac13 t (x \# y , z) I -  (x \# y)  z + \{\text{cyclic permutations}\}$ is linear and symmetric in $x,y,z$. It is indeed the polarization of \eqref{cubic}, hence it is zero, implying that $\Md = 0$. We stress that it is crucial to have associativity with respect to the standard matrix product of elements in $\jotq$, in order to apply the polarization statement; we do need in particular $x^2 x = x x^2 = x^2 \cdot x$, which does indeed hold in the associative case.

Analogous calculations for the other terms in the r.h.s. of \eqref{jazzz} involving $\{B^+, A^- , c\}$, $\{B^+, A^- , \cu\}$, $\{B^+, A^+ , C^+\}$ plus their cyclic permutations prove that $\jac_{11} = 0$.

Next, we proceed to consider $\jac_{12}$ which, after some algebra, can be computed to read :

\beas{l}
\jac_{12} = \\
\left( [a,b] \otimes I + I \otimes [\au,\bu] + A^+ \diamond B^- - B^+ \diamond A^- \right) C^+ \\
-\left( (a \otimes I) B^+ -  (b \otimes I) A^+ + (I \otimes \au) B^+ + B^+ (I \otimes \aud) \right.\\
\left. - (I \otimes \bu) A^+ - A^+ (I \otimes \bud) +  A^- \times B^- \right) (I\otimes \cud) \\
- (c\otimes I + I \otimes \cu) \left( (a \otimes I) B^+ -  (b \otimes I) A^+ + (I \otimes \au) B^+ + B^+ (I \otimes \aud) \right.\\
\left. - (I \otimes \bu) A^+ - A^+ (I \otimes \bud) +  A^- \times B^- \right) \\
- C^+ \left( I \otimes [\aud,\bud] + A^- \bullet B^+ - B^- \bullet A^+ \right) \\
+ \left( - B^- (a \otimes I)  +  A^- (b \otimes I) - (I \otimes \aud) B^- - B^- (I \otimes \au) \right. \\
\left. + (I \otimes \bud) A^- + A^- (I \otimes \bu) +  A^+ \times B^+ \right) \times C^- + \text{cyclic permutations}.
\eeas

Many terms cancel out trivially, and one remains with terms of the following three types:

\beas{l}
1) \quad (I \otimes \cu) (A^- \times B^-) + (A^- \times B^-) (I \otimes \cud) \\
+ ((I \otimes \cud) A^-) + (A^- (I \otimes \cu)) \times B^- - ((I \otimes \cud) B^-) + (B^- (I \otimes \cu)) \times A^- \\ \\
2)\quad -(c\otimes I) (A^- \times B^-)  - (A^- (c \otimes I)) \times B^- + (B^- (c \otimes I)) \times A^-\\ \\
3) \quad (A^+ \times B^+) \times C^- + (B^+\diamond C^-) A^+ - (A^+\diamond C^-) B^+ \\
+ A^+(C^-\bullet B^+) - B^+(C^-\bullet A^+),
\eeas

where we remark that the first two terms show the action of the $\af$ and $\ad$ subalgebras as  derivations.

Let us analyze each of the terms 1) - 3) separately.

1) Writing this term explicitly, one obtains:
\beas{l}
\epsilon_{ijk} (\cu (a_j\# b_k) + (a_j\# b_k) \cud) + \epsilon_{ijk} (\cud a_j + a_j \cu) \# b_k - \epsilon_{ikj}(\cud b_k + b_k \cu) \# a_j ) \\
= \epsilon_{ijk} \left[ \cu (a_j\# b_k) + (a_j\# b_k) \cud +(\cud a_j + a_j \cu) \# b_k +(\cud b_k + b_k \cu) \# a_j \right]
\eeas
In order to show that the expression in brackets is identically zero, we write $\sltq \ni \cu = h + s$, namely as the sum of a traceless Hermitian matrix $h$ and of a skew-Hermitian matrix $s$.
Since the expression under consideration is linear in $\cu$, we can consider the two contributions of $h$ and $s$ separately. The contribution of $h$ reads:
\beas{l}
4\left((a\jdot b) \jdot h + (a\jdot h) \jdot b + (b\jdot h) \jdot a \right)- 4 \left( t(a) b\jdot h +  t(b) h\jdot a + t(h) a\jdot b \right)\\
- 2 \left[ \left( t(b,h) - t(b)t(h)\right) a + \left( t(h,a) - t(h)t(a)\right) b + \left( t(a,b) - t(a)t(b)\right) h \right] \\
- 2 \left( t(a,b\jdot h) + t(a\jdot h,b) - t(a,h)t(b) - t(b,h) t(a) - t(a,b)t(h) + t(a)t(b)t(h) \right) I,
\eeas
where we have added all terms in $t(h)$, since $h$ is traceless. By adopting the first identity in \eqref{idmj} we see that we have obtained a symmetric multilinear form that is ($12$ times) the polarization of \eqref{cubic}, hence it is zero.

On the other hand, the contribution of $s$ can be easily shown to be zero, because $sx-xs$ is a derivation for $x\in \jotq$; we have indeed that if $D$ is a derivation $t(Dx)=0$ and
$ t(D(a),b) + t(a,D(b)) = 0$. Hence:
\be
D( a \# b) = 2 D(a) \jdot b + 2 a \jdot D(b) - D(a) t(b) - D(b) t(a) = D(a) \# b + a \# D(b),
\ee

and this implies the vanishing of the contribution of $s$ to term reported at point 1.

2) We can write this expression as:
\bes
- (\epsilon_{ijk} c_{\ell i} + \epsilon_{\ell i k} c_{j i} - \epsilon_{\ell ij} c_{k i})(a_j \# b_k)
\ees
For $\ell = k$, or $\ell = j$, or $k=j$, the first round bracket trivially vanishes. For $\ell \ne j \ne k \ne \ell$, it can be written as
$\epsilon_{\ell jk} c_{\ell \ell} + \epsilon_{\ell jk} c_{j j} + \epsilon_{\ell jk} c_{k k} = 0$, since $t(c) = 0$.

3) Explicit calculation shows that the $i$-th component of this term reads:
\beas{l}
(a_\ell \# b_i)\# c_\ell - (a_i \# b_\ell)\# c_\ell + t(b_\ell , c_\ell) a_i + t(a_i , c_\ell) b_\ell - t(a_\ell , c_\ell) b_i - t(b_i , c_\ell) a_\ell \\
- a_i c_\ell b_\ell - b_\ell c_\ell a_i  + a_\ell c_\ell b_i + b_i c_\ell a_\ell \\
=- \{ a_\ell , c_\ell , b_i \} + \{ a_i , c_\ell , b_\ell \} - a_i c_\ell b_\ell - b_\ell c_\ell a_i  + a_\ell c_\ell b_i + b_i c_\ell a_\ell = 0,
\eeas
where the triple product $\{ x , y , z \} := V_{x,y}z$ has been introduced in section \ref{sec:jp} and, in the associative case we are considering here: $\{ x , y , z \} = xyz + zyx$, thus implying that also the term 3) vanishes.\\

This ends the proof of the fact that $\jac_{12} = 0$.

Analogous calculations show that also $\jac_{21}= \jac_{22} = 0$, thus proving the Jacobi identity for the matrix realization $\rep$ \eqref{mest} of the adjoint of $\est$, implying the Jacobi identity for the matrix realizations $\rep$ \eqref{mfq} and \eqref{mes} of the adjoint of $\fq$ and $\es$, respectively. $\blacksquare $


\section{$\mathbf{n}=8$ : Matrix representation of $\eo$}\label{sec:e8}

Finally, we consider the case of $\eo$, the largest finite-dimensional exceptional Lie algebra.

We use the notation $L_x z := x\jdot z$ and, for $\vx \in \cc^3 \otimes \joto$ with components $(x_1, x_2, x_3)$, $L_\vx \in \cc^3 \otimes L_{\joto}$ denotes the corresponding operator-valued vector with components $(L_{x_1}, L_{x_2}, L_{x_3})$.

We can write an element $\au$ of $\es$ as $\au =  L_x + \sum [L_{x_i},L_{y_i}]$ where $x,x_i,y_i \in \joto$ ($i=1,2,3$) and $t(x) = 0$, \cite{schafer2} \cite{jacob2}. The adjoint is defined by $\aud:= L_x - [L_{x_1},L_{x_2}]$. Notice that the operators $F := [L_{x_i},L_{y_i}]$ span the $\fq$ subalgebra of $\es$, namely the derivation algebra of $\joto$  (recall that the Lie algebra of the structure group of $\joto$ is $\es \oplus \cc$).

We should remark that $(\au,-\aud)$ is a derivation in the Jordan Pair $(\joto,\jobto)$, and it is here useful to recall the relationship between the structure group of a Jordan algebra $J$ and the automorphism group of a Jordan Pair $V = (J,J)$ goes as follows \cite{loos1}: if $g \in Str(J)$ then $(g, U^{-1}_{g(I)} g) \in Aut(V)$. In our case, for $g = 1 + \epsilon (L_x + F)$, at first order in $\epsilon$ we get (namely, in the tangent space of the corresponding group manifold) $ U^{-1}_{g(I)} g = 1 + \epsilon (- L_x + F) +O(\epsilon^2)$.

Next, we introduce a product $\mep$,\cite{bt-2}, such that $L_x \mep L_y :=  L_{x\cdot y} + [L_x, L_y]$, $F\mep L_x := 2 F L_x$ and $L_x \mep F :=2  L_x F$ for $x,y \in \joto$, including each component $x$ of $\vx \in \cc^3\otimes \joto$ and $y$ of $\vy \in \cc^3\otimes \joto$. By denoting with $[\ ;\ ]$ the commutator with respect to the $\mep$ product, we also require that $[F_1 ; F_2] := 2 [F_1,F_2]$. One thus obtains that $L_x \mep L_y + L_y \mep L_x = 2 L_{x\cdot y}$ and $[F; L_x] := F\mep L_x - L_x \mep F= 2 [F, L_x] = 2 L_{F(x)}$, where he last equality holds because $F$ is a derivation in $\joto$.\\

Therefore, for $\fqe \in \eo$, we write:
\begin{equation}
\rep(\fqe) = \left ( \begin{array}{cc} a\otimes Id + I \otimes \au & \Lvxp \\ \Lvxm & -I\otimes \aud
\end{array}\right)
\label{meo}
\end{equation}
where $a\in \adc$, $\au \in \es$, and we recall that $I$ is the $3\times 3$ identity matrix, as above; furthermore, $Id := L_I$ is the identity operator in $L_{\joto}$ (namely, $L_I L_x= L_x$). Notice that $Id$ is the identity also with respect to the $\mep$ product.

By extending the $\mep$ product in an obvious way to the matrix elements \eqref{meo}, one achieves that $(I \otimes \au) \mep \Lvyp + \Lvyp \mep (I \otimes \aud) = 2 L_{(I \otimes \au) \vyp}$ and  $(I \otimes \aud) \mep \Lvym + \Lvym \mep (I \otimes \au) = 2 L_{(I \otimes \aud) \vym}$.

After some algebra, the commutator of two matrices like \eqref{meo} can be computed to read :
\be
\begin{array}{c}
\left[
\left ( \begin{array}{cc} a\otimes Id + I \otimes \au & \Lvxp \\ \Lvxm & -I\otimes \aud
\end{array}\right) ,
\left ( \begin{array}{cc} b\otimes Id + I \otimes \bu &\Lvyp \\ \Lvym & -I\otimes \bud
\end{array}\right) \right] \\  \\ :=
\left (\begin{array}{cc} C_{11} & C_{12}\\
C_{21} & C_{22}
 \end{array} \right) \hfill
\label{eocom}
\end{array}
\ee
where:
\be
\begin{array}{ll}
C_{11} &= [a,b] \otimes Id + 2 I \otimes [\au,\bu] + \Lvxp \diamond \Lvym - \Lvyp \diamond \Lvxm \\ \\
C_{12} &=  (a \otimes Id) \Lvyp -  (b \otimes Id) \Lvxp +2  L_{(I \otimes \au) \vyp}\\
 &\phantom{:=} - 2 L_{(I \otimes \bu) \vxp} +  \Lvxm \times \Lvym \\ \\
C_{21} &= - \Lvym (a \otimes Id)  +  \Lvxm (b \otimes Id) - 2 L_{(I \otimes \aud) \vym} \\
&\phantom{:=} +2  L_{(I \otimes \bud) \vxm} +  \Lvxp \times \Lvyp \\ \\
C_{22} &= 2 I \otimes [\aud,\bud] + \Lvxm \bullet \Lvyp - \Lvym \bullet \Lvxp.
\end{array}
 \label{comreleo}
\ee
It should be stressed that the products occurring in \eqref{comreleo} do differ from those of  \eqref{not1}; namely, they are defined as follows\footnote{It should be stressed here that the matrix products $x\diamond y$, $x\cdot y$
e $x\times y$ defined in (\ref{not1eo}), never appeared (to the best of our
present knowledge) in the literature, and are an original result of the
present investigation.} :
\be
\begin{array}{ll}
\Lvxp \diamond \Lvym &:= \left(\frac13 t(x^+_i, y^-_i) I - t(x^+_i,y^-_j) E_{ij}\right) \otimes Id +\\
&\phantom{:=} I \otimes \left(\frac13 t(x^+_i, y^-_i ) Id - L_{x^+_i \cdot y^-_i} - [L_{x^+_i}, L_{y^-_i}] \right) \\ \\
\Lvxm \bullet \Lvyp &:= I \otimes (\frac13 t(x^-_i,y^+_i) Id - L_{x^-_i \cdot y^+_i} - [L_{x^-_i}, L_{y^+_i}]) \\  \\
\Lvxpm \times \Lvypm &:= L_{\vxpm \times \vypm} = L_{\epsilon_{ijk} (x_j^\pm \# y_k^\pm)}.
\end{array}
 \label{not1eo}
\ee

From the properties of the triple product of Jordan algebras (discussed in Sec. 2), it holds that $ L_{x^+_i \cdot y^-_i} + [L_{x^+_i}, L_{y^-_i}] = \frac12 V_{x^+_i , y^-_i} \in \es\oplus \cc$, see \eqref{vid}. Moreover, one can readily check that $[a_1^\dagger,b_1^\dagger] = - [a_1,b_1]^\dagger$, $(a\otimes Id) L_b = L_{(a\otimes Id) b}$ and $\Lvym \bullet \Lvxp = I \otimes (\dfrac13 t(x^+_i,y^-_i) Id - L_{x^+_i \cdot y^-_i} - [L_{x^+_i}, L_{y^-_i}])^\dagger$; this result implies that we are actually considering an algebra.

In the next section we are going to prove that Jacobi's identity holds for the algebra of Zorn-type matrices \eqref{meo}, with Lie product given by \eqref{eocom} - \eqref{not1eo}. On the other hand, once Jacobi's identity is proven, the fact that the Lie algebra so represented is $\eo$ is made obvious by a comparison with the root diagram in figure \ref{fig:diagram}, for $n=8$; in this case, we have:
\begin{itemize}
\item[1)] an $g_{0}^{8}=\es$, commuting with $\adc$;
\item[2)] As in general, the three Jordan Pairs which globally transform as a $(\mathbf{3},\overline{\mathbf{3}})$ of $\adc$; in this case, each of them transforms as a $(\mathbf{27}, \overline{\mathbf{27}})$ of $\es$.
\end{itemize}

As a consequence, we reproduce the well known branching rule of the adjoint of $\eo$ with respect to its maximal and non-symmetric subalgebra $\adc \oplus \es $:
\begin{equation}
\mathbf{248}=\left( \mathbf{8},\mathbf{1}\right) +\left( \mathbf{1},\mathbf{%
78}\right) +\left( \mathbf{3},\mathbf{27}\right) +\left( \overline{\mathbf{3}%
},\overline{\mathbf{27}}\right) .
\end{equation}


\section{Jacobi identity for $\eo$}\label{sec:jacobieo}

We use the same notation as in section \ref{sec:jacobi}, and write \eqref{meo} in a slight different way, namely, for for $\fqe_1 \in \eo$:
\begin{equation}
\rep(\fqe_1) = \left ( \begin{array}{cc} a\otimes I + I \otimes \au & A^+ \\ A^- & -I\otimes \aud
\end{array}\right),
\label{meo1}
\end{equation}
where $a\in \adc , \au \in\es$ and $A^+ , A^-$ three vectors with elements in $\joto , \jobto$. Similarly, one can define $\rep(\fqe_2)$ and $\rep(\fqe_3)$ by respectively replacing $a \to b$ and $a\to c$ in \eqref{meo1}. Let us then write:
\be
[[\rep(\fqe_1),\rep(\fqe_2)],\rep(\fqe_3)]]+\text{cyclic permutations}
:= \left (\begin{array}{cc} \jac_{11} & \jac_{12}\\
\jac_{21} & \jac_{22}
 \end{array} \right)
\ee

In order for the Jacobi identity to hold for the matrix realization \eqref{meo1} of the adjoint of $\eo$, we have to prove that $\jac_{11} =\jac_{12} =\jac_{21}=\jac_{22} =0$.

After some algebra, we compute:
\beas{l}
 \jac_{11} =\\
\  [[a,b],c] \otimes Id + 4 I \otimes [[\au,\bu],\cu] + (A^+ \diamond B^- - B^+ \diamond A^-)(c\otimes Id + I\otimes \cu) \\
 - (c\otimes Id + I\otimes \cu)(A^+ \diamond B^- - B^+ \diamond A^-) \\
+ \left( (a \otimes Id) B^+ -  (b \otimes Id) A^+ + (I \otimes \au) B^+ + B^+ (I \otimes \aud) \right.\\
\left. - (I \otimes \bu) A^+ - A^+ (I \otimes \bud) +  A^- \times B^- \right) \diamond C^- \\
- C^+ \diamond \left(   - B^- (a \otimes Id)  +  A^- (b \otimes Id) - (I \otimes \aud) B^- - B^- (I \otimes \au) \right. \\
\left. + (I \otimes \bud) A^- + A^- (I \otimes \bu) +  A^+ \times B^+ \right) + \text{cyclic permutations}
\label{blues}

\eeas

The first two terms in the r.h.s. of \eqref{blues} vanish upon cyclic permutations, because of the Jacobi identity in $\adc$ and $\es$. The terms containing  $A^+ , B^- , c$ can be proved to vanish, by the very same arguments used in section \ref{sec:jacobi}.

Next, we consider the terms containing  $A^+ , B^- , c_1$. By denoting with $a_k , b_k \in \joto$ ($k  = 1,2,3$), the components of $A^+$ and $B^-$ respectively, and using the shorthand notation: $E(x,y):=L_{x\cdot y} + [L_x , L_y] = \frac12 V_{x,y}$, for $x,y\in \joto$, one can compute that:
\beas{l}
[A^+ \diamond B^-;I\otimes \cu]  + ((I\otimes \cu) A^+  + A^+ (I\otimes \cud))\diamond B^- \\
- A^+ \diamond ((I\otimes \cud) B^-  + B^-(I\otimes \cu))\\
= 2 I\otimes [\cu, E(a_i, b_i)] + 2 \left( \frac13 t(\cu (a_i), b_i) I - t(\cu(a_i) , b_j ) E_{ij}\right) \otimes Id \\
\pu + 2 I\otimes \left( \frac13 t(\cu (a_i), b_i) Id - E(\cu(a_i),b_i) \right) \\
\pu -  2\left( \frac13 t(a_i , \cud(b_i)) I -t(a_i , \cud(b_j)) E_{ij} \right) \otimes Id \\
\pu - 2 I \otimes \left( \frac13 t(a_i , \cud(b_i)) Id - E(a_i,\cud(b_i)) \right)\\
= \left( \frac43(t(\cu(a_i), b_i) - t(a_i , \cud(b_i)))  I- 2(t(\cu(a_i), b_j) - t(a_i , \cud(b_j)))E_{ij} \right)\otimes Id \\
+ 2 I \otimes \left(  [\cu,E(a_i,b_i)] - E(\cu(a_i),b_i) + E(a_i,\cud(b_i))  \right).
\label{blues2}
\eeas
In order to prove that \eqref{blues2} sums up to zero, we start and observe that $t(\cu(a),b) = t(a, \cud(b))$; this is easily shown by writing $\cu = L_x + F$ (hence $\cud = L_x - F$) and noticing that $t(L_x a, b) = t(x\cdot a, b) = t(a,x\cdot b) = t(a,L_x(b))$ and $t(F x, y) + t(x, F y) = 0$, being $F$ is a derivation in $\joto$.
Moreover, $[c_1,E(a,b)] = E(\cu(a),b) - E(a,\cud(b))$, by \eqref{dib}.
This indeed implies that \eqref{blues2} vanishes.

Finally, we consider terms in the r.h.s. of \eqref{blues} which contain structures like $(A^- \times B^-) \diamond C^-$; they read:
\bea{l}
(A^- \times B^-) \diamond C^- + (B^- \times C^-) \diamond A^- + (C^- \times A^-) \diamond B^-\\
= \epsilon_{i \ell k} \left( \frac13 t (a_\ell \# b_k , c_i) I - t (a_\ell \# b_k , c_j)E_{ij} \right) \otimes Id \\
\pu + I\otimes \epsilon_{i \ell k} \left( \frac13 t (a_\ell \# b_k , c_i) Id -  E(a_\ell \# b_k,c_i) \right) + \text{cyclic permutations}\\
:= \Mu \otimes I + I \otimes \Md.
\label{apbmc4}
\eea

$\Mu = 0$, by the same argument used in section \ref{sec:jacobi}. Let us here show that $\Md = 0$. In order to do this, we write $(a_j,b_k,c_i) := \frac13 t(a_j\# b_k, c_i)I - (a_j\#b_k)\cdot c_i$. Thence:
\be
\Md = \epsilon_{ijk} \left( L_{(a_j,b_k,c_i)} - [L_{a_j\# b_k} , L_{c_i} ] \right) + \text{cyclic permutations}
\ee
For each fixed i,j,k, it holds that
\bea{l}
 \epsilon_{ijk} (a_j,b_k,c_i) +  \epsilon_{jki} (b_k,c_i,a_j) +  \epsilon_{kij} (c_i, a_j,b_k) \\
= \epsilon_{ijk} \left( (a_j,b_k,c_i) +  (b_k,c_i,a_j) +  (c_i, a_j,b_k) \right)
\eea

Since $(x,y,z)$ is symmetric in $x,y$ and linear in $x,y,z$, the above expression is linear and symmetric in $a_j,b_k,c_i$, thus it is the polarization of $(x,x,x) = 2 ( \frac13 t(\xs , x) I - \xs \cdot x) = 0$, by \eqref{cubic}. Similarly, for $[L_{a_j\# b_k} , L_{c_i} ] $ + cyclic permutations, we get the polarization of $[L_{\xs} , L_x ] $, which is zero by the Jordan identity \eqref{pass}, namely:
\be
[L_{\xs} , L_x ] z= \xs \jdot (x\jdot z) - x \jdot (\xs \jdot z) = x^2 \jdot (x\jdot z) - x \jdot (x^2 \jdot z) = 0 \ \forall z \in J
\ee

Analogous calculations for terms in the r.h.s. of \eqref{blues} which contain structures like $\{B^+, A^- , c\}$, $\{B^+, A^- , \cu\}$, $\{B^+, A^+ , C^+\}$ (plus their cyclic permutations) prove that $\jac_{11} = 0$.

Next, we proceed to consider $\jac_{12}$ which, after some algebra, can be computed to read :

\beas{l}
\jac_{12} = \\
\left( [a,b] \otimes I + 2 I \otimes [\au,\bu] + A^+ \diamond B^- - B^+ \diamond A^- \right) C^+ \\
-\left( (a \otimes I) B^+ -  (b \otimes I) A^+ + (I \otimes \au) B^+ + B^+ (I \otimes \aud) \right.\\
\left. - (I \otimes \bu) A^+ - A^+ (I \otimes \bud) +  A^- \times B^- \right) (I\otimes \cud) \\
- (c\otimes I + I \otimes \cu) \left( (a \otimes I) B^+ -  (b \otimes I) A^+ + (I \otimes \au) B^+ + B^+ (I \otimes \aud) \right.\\
\left. - (I \otimes \bu) A^+ - A^+ (I \otimes \bud) +  A^- \times B^- \right) \\
- C^+ \left( I \otimes [\aud,\bud] + A^- \bullet B^+ - B^- \bullet A^+ \right) \\
+ \left( - B^- (a \otimes I)  +  A^- (b \otimes I) - (I \otimes \aud) B^- - B^- (I \otimes \au) \right. \\
\left. + (I \otimes \bud) A^- + A^- (I \otimes \bu) +  A^+ \times B^+ \right) \times C^- + \text{cyclic permutations}
\eeas

By noticing that $[a_1^\dagger,b_1^\dagger] = - [a_1,b_1]^\dagger$ and that, as already noticed, $(a\otimes Id) L_b = L_{(a\otimes Id) b}$, as already noticed, one finds that many terms cancel out trivially, and only terms of the following three types remain:

\beas{l}
1) \quad -2 L_{(I \otimes \cu) a^- \times b^-} - L_{(I \otimes \cud) a^- \times b^-} + L_{(I \otimes \cud) b^- \times a^- }\\ \\
2)\quad -(c\otimes I) L_{a^- \times b^-}  - L_{a^- (c \otimes I)) \times b^-} + L_{b^- (c \otimes I)\times a^-}\\ \\
3) \quad L_{(a^+ \times b^+) \times c^-} + t(b_i^+,c_i^-)L_{a^+} - t(b_i^+,c_j^-)E_{ij}\otimes Id L_{a^+}  \\
-t(a_i^+,c_i^-)L_{b^+} + t(a_i^+,c_j^-)E_{ij}\otimes Id L_{b^+} - L_{V_{b_i^+, c_i^+} a^+} + L_{V_{a_i^+, c_i^+} b^+}
\eeas

Terms like $1)$ and $2)$ can be shown to vanish using similar arguments to those of section \ref{sec:jacobi}. The $i$-th component of terms like $3)$ can be written as (omitting the $+,-$ superscripts):
\bea{l}
L_{(a_\ell \# b_i) \# c_\ell} -  L_{(a_i \# b_\ell) \# c_\ell} + t(b_\ell,c_\ell)L_{a_i} + t(a_i, c_\ell) L_{b_\ell} \\
-t(a_\ell,c_\ell)L_{b_i} - t(b_i, c_\ell) L_{a_\ell} - L_{V_{b_\ell, c_\ell} a_i} + L_{V_{a_\ell, c_\ell} b_i},
\eea
which vanishes because of \eqref{vid}.

This ends the proof of the fact that $\jac_{12} = 0$.

Analogous calculations show that also $\jac_{21}= \jac_{22} = 0$, thus proving the Jacobi identity for the matrix realization \eqref{meo1} (or, equivalently \eqref{meo}) of the adjoint of $\eo$. $\blacksquare $


\section{Future developments}

There are several topics that we are planning to develop in the future.

One is the extension of the Zorn-type representations to the Lie algebra of
the semi-direct product group $\esm$, through a representation of the
\textit{sextonions} \cite{Westbury,LM-E7-1/2} and of the algebra of their
derivations.

A second interesting venue of developments is the
characterization of all real forms of these representations of the
exceptional Lie algebras, as well as the treatment of \textit{split} forms of Hurwitz's algebras $\mathbf{C}$, $%
\mathbf{Q}$, $\mathfrak{C}$, with a particular attention to the coset spaces
related to the scalar manifolds in supergravity. This  would yield a \textit{Zorn-like}
realization of (some of) the maximal non-symmetric embeddings considered in
\cite{Jordan-Pairs-D=5}, and proved in a broader framework in \cite%
{super-Ehlers-1}.

Moreover, it would be interesting to consider Jordan pairs for \textit{%
semi-simple} Jordan algebras of rank $3$ of relevance for
supergravity theories, along the lines of the treatment given in \cite%
{Jordan-Pairs-D=5}.

We plan then to proceed to the study of the representations of quantum
exceptional groups - in particular \textit{quantum} $\eo$ - and of integrable models
built on them. We aim at a new perspective of elementary particle physics at
the early stages of the Universe based on the idea that interactions, defined in a purely algebraic way, are the fundamental objects of the theory,
whereas space-time, hence gravity, are derived structures.

\section*{Acknowledgments}

The work of AM is supported in part by the FWO - Vlaanderen, Project No. G.0651.11, and in part by the
Interuniversity Attraction Poles Programme initiated by the Belgian Science Policy (P7/37).

The work of PT is supported in part by the \textit{Istituto Nazionale di Fisica Nucleare} grant In. Spec. GE 41.


\end{document}